# In-plane staging in lithium-ion intercalation of bilayer graphene

Thomas Astles[1], James G. McHugh[1,2], Rui Zhang[1], Qian Guo[1,2], Madeleine Howe[1], Zefei Wu[1,2], Kornelia Indykiewicz[1,2], Alex Summerfield[2], Zachary A.H. Goodwin[1,2], Sergey Slizovskiy[1,2], Daniil Domaretskiy[1], Andre K. Geim[1,2], Vladimir Falko[1,2], Irina V. Grigorieva[1,2]

[1] Department of Physics and Astronomy, University of Manchester
[2] National Graphene Institute, University of Manchester

**The ongoing efforts to optimize Li-ion batteries led to the interest in intercalation of nanoscale layered compounds, including bilayer graphene. Its lithium intercalation has been demonstrated recently but the mechanisms underpinning the storage capacity remain poorly understood. Here, using magnetotransport measurements, we report *in-operando* intercalation dynamics of bilayer graphene. Unexpectedly, we find four distinct intercalation stages that correspond to well-defined Li-ion densities. We refer to these stages as 'in-plane', with no in-plane analogues in bulk graphite. The fully intercalated bilayers represent a stoichiometric compound $C_{14}LiC_{14}$ with a Li density of $\sim 2.7 \cdot 10^{14}$ cm$^{-2}$, notably lower than fully intercalated graphite. Combining the experimental findings and DFT calculations, we show that the critical step in bilayer intercalation is a transition from AB to AA stacking which occurs at a density of $\sim 0.9 \cdot 10^{14}$ cm$^{-2}$. Our findings reveal the mechanism and limits for electrochemical intercalation of bilayer graphene and suggest possible avenues for increasing the Li storage capacity.**

## Introduction

Graphite as an anode material for Li-ion batteries has been studied extensively for many decades and is currently being employed as an essential component in rechargeable batteries [1-4]. Often used in the form of graphite powder baked onto copper foils, it offers many advantages for battery anodes such as chemical inertness, reversible intercalation, good cyclability and relatively low costs. In an effort to further enhance the performance of graphite-based anodes, recent research focused on replacing graphite with few-layer graphene and attempting to establish the factors that govern intercalation of Li ions, particularly into BLG that presents the elementary building block of the AB-stacked (Bernal) graphite [5-17]. Promisingly, Li diffusion rates in BLG are found to be two orders of magnitude higher than in bulk graphite [5], which may provide much faster charging-discharging. On the other hand, the storage capacity for BLG – Li-ion density achievable by electrochemical intercalation – has so far been disappointingly low ($n_{Li} \lesssim 2 \times 10^{14}$ cm$^{-2}$ [5,9,10]), that is, 3 times lower than that achieved for stage-I intercalation of bulk graphite ($LiC_6$) [18-21]. The only exception is a recent study where intercalation was monitored in a transmission electron microscope (TEM), which reported formation of multilayered metallic lithium [7]. In the latter case, however, the high-energy electron beam might interfere with the electrochemical process by, for example, contributing to Li ions reduction. The question remains open whether Li-ion storage above $\sim 2 \times 10^{14}$ cm$^{-2}$ can be achieved in operando, using the standard electrochemistry most relevant for battery technologies.

Understanding the limits for the use of bilayer and few-layer graphene for Li-ion storage requires knowledge of mechanisms governing Li intercalation, including any structural changes, arrangements of Li ions relative to the underlying graphene lattice, and effects of disorder. While the above factors have been the focus of many computational studies [11,13-17,22], experimental evidence remains scarce. Among other outstanding questions is the possibility of in-plane staging, that is, of a sequence of preferential Li configurations replacing each other during intercalation, by analogy with alkali ion intercalation of bulk graphite [23-27]. Such staging has been suggested to explain certain features of Li intercalation of a 'bilayer foam', a bulk system consisting predominantly (but not exclusively) of graphene bilayers [8]. However, no signatures of in-plane staging were reported in several other studies of isolated BLG [5,6,9], implying that the features observed in ref. 8 could have a different origin (e.g., staging in graphene multilayers present in that system, or elastic strain [18,28]).

To shed light on the above uncertainties, in this work we study Li intercalation of BLG using an on-chip electrochemical cell where the entry and exit of Li ions are examined via changes in resistivity and carrier density of graphene. The intercalation process is continuously monitored over many intercalation-deintercalation cycles, while the rate of the intercalation reaction is kept sufficiently slow to resolve individual stages. We observe four distinct plateaus in graphene resistivity, corresponding to distinct densities of intercalated lithium, $n_{\text{Li}}$. We refer to these as in-plane stages (I to IV), not to confuse the effect with the well-known staging in bulk graphite, where Li ions fill interlayer spaces to the full capacity in one step. Lithium intercalation in our device setup is further verified by *in operando* Raman spectroscopy. Our measurements of Li ion densities for stages III and IV allow us to identify them as $C_{18}LiC_{18}$ and $C_{14}LiC_{14}$, where Li ions attain hexagonal arrangements commensurable with the underlying graphene lattice. Assuming similar commensurability that minimizes Coulomb interactions, low-density stages II and I are identified as $C_{38}LiC_{38}$ and $C_{42}LiC_{42}$. Combining the experimental findings and DFT calculations, we show that $C_{14}LiC_{14}$ corresponds to the thermodynamic equilibrium for intercalated AA-stacked bilayers. Another observed stage ($C_{18}LiC_{18}$) is close to this equilibrium configuration, whereas stages I and II at much lower $n_{\text{Li}}$ are attributed to metastable states within the original AB stacking. Transitions between the stages occur rapidly (typically within 1 sec) over the entire device area. Our DFT analysis suggests that the transition between the two pairs of in-plane staging is accompanied by changing the BLG structure from AB to AA stacking, which occurs at a threshold Li ion density in the AB bilayer and is facilitated by formation of AB/BA boundaries during intercalation-deintercalation cycles.

## Results

A schematic of the experimental setup and an optical image of one of our devices are shown in Fig. 1a&b. BLG was mechanically exfoliated onto a Si/SiO$_2$ substrate and shaped into a Hall bar, with a few μm of its edge being exposed to solid polymer electrolyte LiTFSI-PEO (Supplementary Methods 1.1&1.2). A gate voltage $V_g$ between the Pt counter electrode and BLG provided a controlled driving force for electrochemical intercalation. To protect the device from degradation, most of the Hall bar and Au contacts were covered with a passivating layer of SU-8 resist as shown in Fig. 1b and Supplementary Fig. 1a. This design ensured that Li ions from the electrolyte could enter BLG's interlayer space (gallery) only through the exposed edge. All measurements were performed in the inert environment of a glovebox with <0.5 ppm oxygen and moisture levels to prevent degradation of the electrolyte. The temperature was kept at 330±2 K to ensure sufficient electrolyte conductivity (Supplementary Fig. 1c,d). Intercalation and deintercalation were monitored as a function of time *t* at 1s intervals via measurements of graphene's resistivity $\rho_{xx}$ and Hall voltage $V_{xy}$ using the standard lock-in technique. Employing different pairs of contacts allowed us to probe the uniformity of intercalation over the ~20 μm length of our bilayer devices (Fig. 1b). If we swept $V_g$, a sharp peak appeared in $\rho_{xx}$ and a spike-like feature in $V_{xy}$ at a critical value of about -3 V (Fig. 1c, Supplementary Fig. 2), which indicated that our initially p-doped devices (Supplementary Methods 1.3) changed their doping polarity and the Fermi level passed through the neutrality point as the result of Li ions entering BLG [5,9]. In measurements of $\rho_{xx}(t)$ and $V_{xy}(t)$, we mostly used $V_g$ = -7V. Smaller $V_g$ required longer times to achieve the same intercalation level but did not affect the discussed results, which is consistent with the known behavior for Li intercalation of graphite [29]. Further details explaining the working of our electrochemical cell are provided in Supplementary Methods 1.3. Below, we focus on the behavior exhibited by one of our devices that was studied in greater detail (device A), and another device B is described in Supplementary Methods (Supplementary Fig. 4). The reported in-plane staging was observed in all five studied BLG devices, with not only qualitatively but also quantitatively the same characteristics (Supplementary Fig. 6).

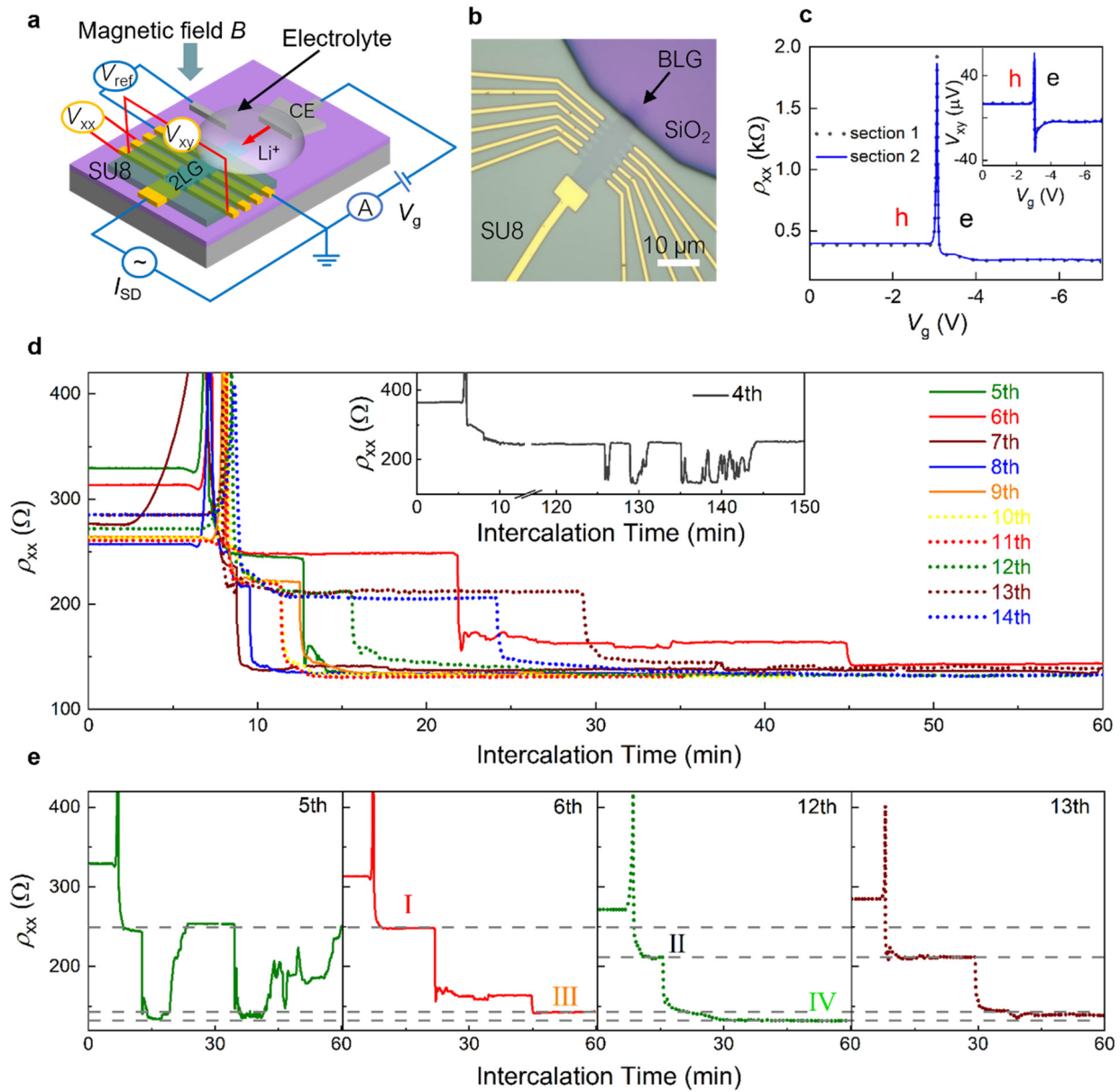

**Figure 1 | Device design and the evolution of BLG resistance over multiple intercalation cycles.** (a) Schematic of the experimental setup. (b) Optical image of device A. (c) Longitudinal resistance (main panel) and Hall voltage (inset) vs the gate voltage. Solid blue and dotted black curves, corresponding to different sections of the device, overlap. (d) Evolution of $\rho_{xx}$ during the first hour in consecutive intercalation cycles (measured at 1s intervals; color coded). The inset shows multiple jumps in $\rho_{xx}$ observed in the 4$^{th}$ intercalation cycle between resistance values corresponding to transitions between stages I and IV (see text). (e) Detailed time evolution for several representative intercalation cycles. Dashed horizontal lines indicate the average values of $\rho_{xx}$ for the found four stages (see text). Infrequently, we also saw plateaus that were not reproduced in any other cycle (see, e.g., the plateau at about 160 Ω in the 6$^{th}$ cycle).

Typical evolution of BLG's resistance over consecutive intercalation cycles is shown in Fig. 1d with representative cycles shown separately in Fig. 1e. The most notable feature here is the presence of resistance plateaus with well-defined $\rho_{xx}$ values that persisted over extended time intervals up to 20 min and, importantly, reappeared in different cycles and for all the studied devices. Only the time span of the plateaus varied from cycle to cycle (Fig. 2c). Transitions between the plateaus occurred typically within 1 sec (our time resolution) and simultaneously across the whole device, see Fig. 2a for a typical example. Studying different cycles and devices, we found that $\rho_{xx}$ for the four observed plateaus was on average ~248, 219, 142 and 134 $\Omega$ with a standard deviation of ±3% (Fig. 1d&e, Supplementary Fig. 4a&b). Plateaus in $\rho_{xx}$ were accompanied by plateaus in $V_{xy}$ measured simultaneously (Supplementary Fig. 3b). To determine the Hall resistivity $\rho_{xy}$ at the plateaus, we reversed the magnetic field (switching it between ±330 mT by rotating a permanent magnet around the intercalation/measurement setup within the glovebox). This allowed us to avoid spurious offsets in $V_{xy}$ which sometimes appeared from a $\rho_{xx}$ contribution. From the $\rho_{xy}$ value, we found the electron density $n$ induced in BLG and then estimated $n_{Li}$ using the known charge transfer of ~0.9$e$ from each Li ion to graphene [11,14,15] ($e$ is the electron charge), see Supplementary Methods 1.3 for details. The Li-ion density for the observed plateaus was ~0.9, 1.0, 2.1 and 2.7 ×10$^{14}$ cm$^{-2}$ (Fig. 3a). In the following, we refer to these four distinct states of BLG intercalation as in-plane stages I to IV, respectively.

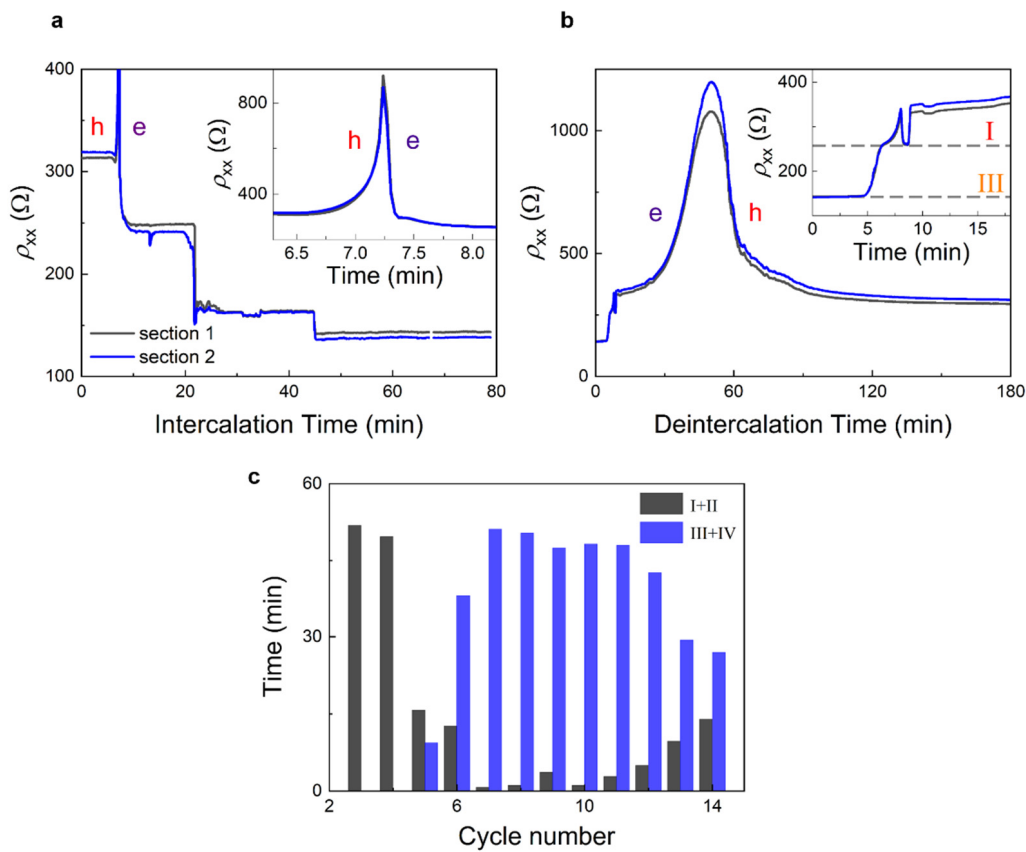

**Figure 2 | Simultaneous changes and kinetics of intercalation stages. (a,b)** Representative time evolution of $\rho_{xx}$ for two sections of device A during intercalation (a) and deintercalation (b); 6$^{th}$ intercalation-deintercalation cycle. Inset in (a): zoom-in of the resistance peak corresponding to the change from p- to n-doping as Li ions enter BLG. The inset in (b) shows typical kinks in $\rho_{xx}$ during deintercalation. Dashed lines in (b) correspond to 258 $\Omega$ and 143 $\Omega$, resistance values characteristic of stage I and stage III intercalation, respectively. **(c)** Overall time spent at stages I and II (AB stacking) and stages III and IV (AA stacking) during consecutive 60-min-long intercalation cycles.

To achieve the reproducible in-plane staging required several cycles of intercalation/deintercalation. During initial few cycles, $\rho_{xx}$ and $V_{xy}$ fluctuated with time without acquiring any specific values (Supplementary Figs 2a & 3a) despite $V_g$ being kept constant and applied for several hours. Remarkably, even in this weakly-doped state the observed fluctuations accurately reproduced at all voltage contacts (Supplementary Fig. 2), which showed that Li-ion doping occurred practically simultaneously over the entire 20 µm long device and Li ions rearranged themselves into different configurations within 1 sec. This also meant that no macroscopic domains were formed, in contrast to, e.g., Li intercalation of bulk $Li_xCoO_2$ [30]. After the first couple of cycles, stage I with its $\rho_{xx} \approx 248\ \Omega$ and $n_{Li} \approx 0.9 \times 10^{14}$ cm$^{-2}$ clearly developed, persisting for a long time (inset of Fig. 1d; Supplementary Fig. 3a). In the 4th or 5th cycle and after long (>2 h) exposures to $V_g$ = -7V, we also started to observe stage IV (lowest $\rho_{xx} \approx 134\ \Omega$ and highest $n_{Li} \approx 2.7 \times 10^{14}$ cm$^{-2}$; inset of Fig. 1d and left panel of Fig. 1e). The latter plots show sharp jumps between the resistance plateaus corresponding to stages I and IV, but the high-$n_{Li}$ states persisted no longer than a few minutes at a time. Only after further cycles, all four in-plane intercalation stages became established and recurred for extended periods of time (Figs 1d&e). Having said that, in later cycles, stage III became poorly defined (transient), gradually transforming into stage IV (Supplementary Fig. 3c; Fig. 1e). In addition, in these later cycles our devices tended to bypass stage I, immediately entering stage II and then making a two-/three- fold jump (in $\rho_{xx}$ and $n_{Li}$, respectively) into the most-doped state (stage IV). The observed 'initial softening or training' of graphite-based systems during intercalation is well known in the literature and attributed to gradual expansion of the interlayer space during initial intercalation cycles [12,18,21]. Let us also emphasize that no further increase in Li density, beyond $n_{Li} \approx 2.7 \times 10^{14}$ cm$^{-2}$ for stage IV, could be detected in any of our devices, irrespective of the driving potential, intercalation time and other conditions. This doping is in fact comparable or higher than those reported previously for Li intercalation of BLG [5,9]. This suggests that stage IV represents the intrinsic capacity of BLG for the standard electrochemical intercalation with Li.

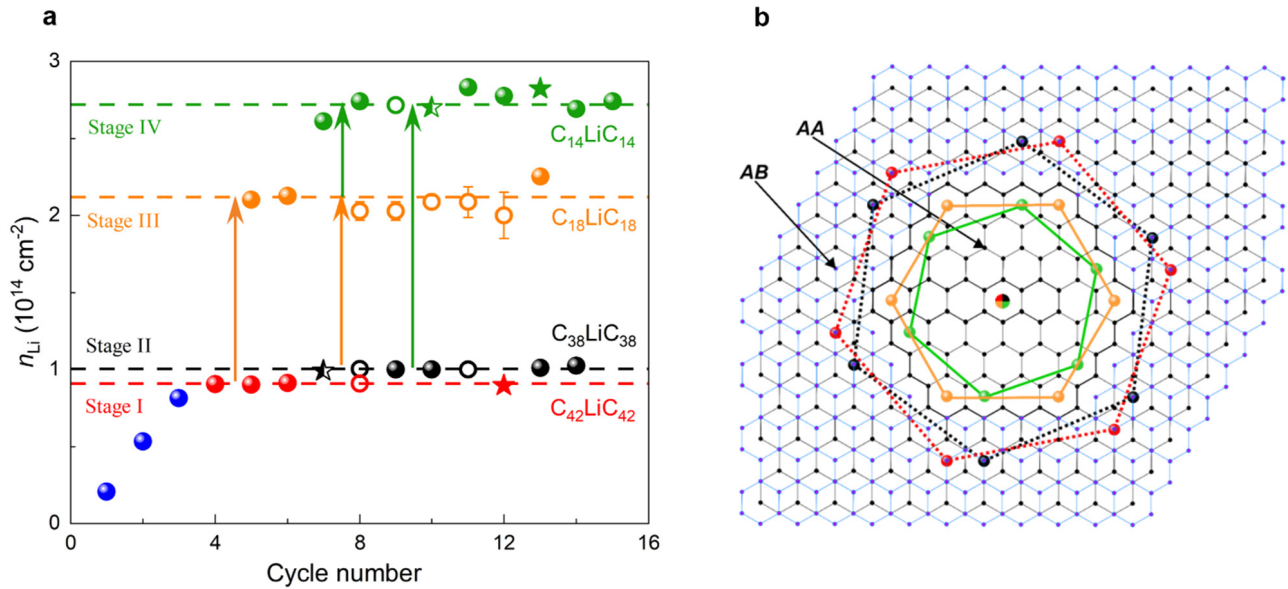

**Figure 3 | Stages of intercalation.** (**a**) Li-ion densities found from Hall measurements (filled symbols). Open symbols show $n_{Li}$ evaluated from measured $\rho_{xx}$ assuming the electron scattering times remained unchanged for a given intercalation stage (Supplementary Methods 1.3). All circular symbols are for device A. For comparison, star symbols show average $n_{Li}$ for corresponding intercalation stages found for two other devices (device B and C). Horizontal lines show $n_{Li}$ expected for exact $C_{42}LiC_{42}$, $C_{38}LiC_{38}$, $C_{18}LiC_{18}$ and $C_{14}LiC_{14}$ stoichiometries (Supplementary Methods 1.3). The arrows indicate dominant transitions that depended on the cycle number. Error bars for open symbols are due to uncertainty in determining $n$ to infer $n_{Li}$. (**b**) Proposed relative positions of Li ions for different stages of intercalation, color-coded as in (a). Stages I and II correspond to intercalation of AB-stacked bilayer, and stages III and IV to AA bilayer. Li

ion positions for AB stacking correspond to one of the two equivalent configurations and are therefore shown by the dashed lines.

As for deintercalation, in contrast to the fast entry of Li ions into the interlayer space, the return to a deintercalated state (after $V_g$ was set back to zero) was much slower and became progressively slower still with each cycle (Fig. 4 and Supplementary Fig. 5). Kinks and small steps indicating the in-plane staging were also observed on $\rho_{xx}(t)$ curves during deintercalation (insets of Figs. 2b and 4) but they were much less pronounced and lasting than those for intercalation half-cycles. Furthermore, unlike the well-defined resistance state reached after each full intercalation to stage IV, $\rho_{xx}$ after deintercalation varied considerably and usually decreased with repeated cycling (see Fig. 1c at $t = 0$) whereas the final carrier density (p-doping) reached after deintercalation remained approximately the same. This is unexpected because disorder induced by intercalation/deintercalation cycles should generally increase the resistivity. Indeed, another notable feature of deintercalation is that the resistance peak at the transition from n- to p- doping became progressively broader and smaller with repeated cycling (Supplementary Fig. 5), which indicated increasing inhomogeneity of deintercalated BLG [31], as expected. Only after many intercalation cycles (typically, >10), disorder started playing a critical role in the measured characteristics. Eventually, after 15-20 cycles, intercalation becomes progressively less effective, and the maximum Li concentrations of $\sim 2.7 \times 10^{14}$ cm$^{-2}$ could no longer be achieved even after many hours (leading to $n_{Li} \ll 2 \times 10^{14}$ cm$^{-2}$).

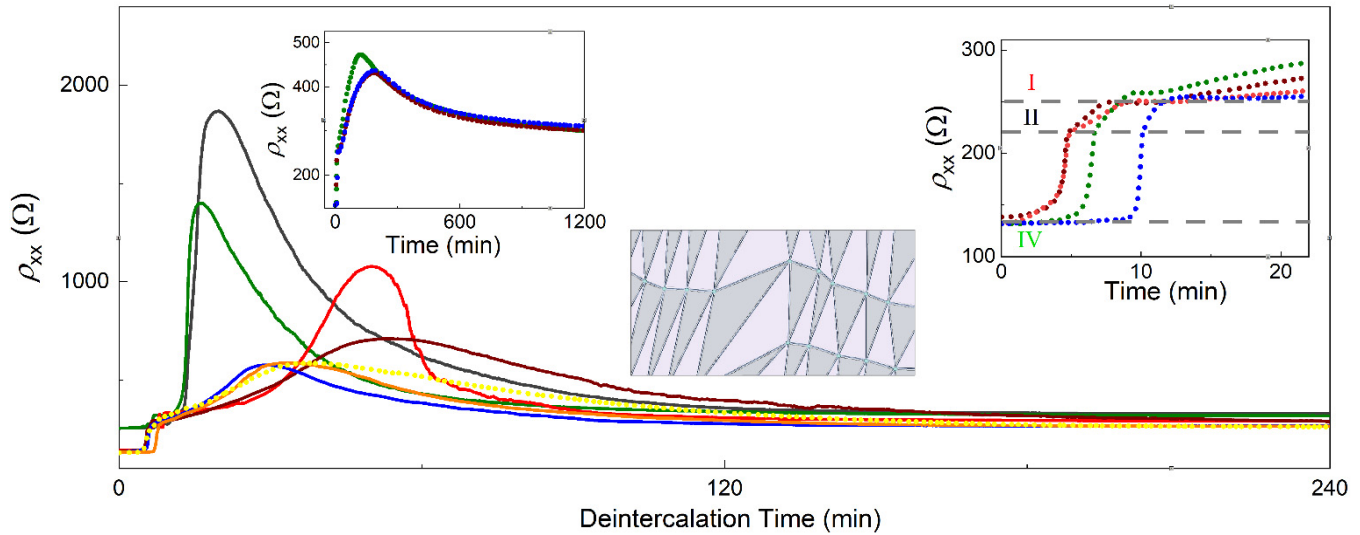

**Figure 4 | Evolution of BLG resistance during deintercalation.** Main panel: $\rho_{xx}(t)$ during the deintercalation part of consecutive cycles (4$^{th}$ to 14$^{th}$, as for intercalation in Fig. 1d, same color coding). All deintercalation curves, including the first three cycles are shown in Supplementary Fig. 5a. The right inset highlights distinct kinks in $\rho_{xx}$ corresponding to transitions between different stages during deintercalation; grey lines correspond to $\rho_{xx}$ = 134, 222 and 250 Ω. The left inset shows evolution of $\rho_{xx}$ during the 12$^{th}$, 14$^{th}$ and 15$^{th}$ deintercalation cycles, in which the fully deintercalated state took considerably longer to reach than the time span in the main panel. Middle inset: schematic of AB/BA domains expected to appear in bilayer graphene under cyclical strain.

As an alternative way to monitor intercalation, we used in situ Raman spectroelectrochemistry (details in Supplementary Methods 1.4). Figure 5 shows the evolution of the Raman G and 2D peaks of our bilayer graphene as a function of the applied gate voltage $V_g$. The G peak position shifted sharply by ~8 cm$^{-1}$ as soon as the gate voltage exceeded the threshold value (3.8V for this device) while the longitudinal resistance (measured simultaneously) went through a sharp peak, indicating intercalation. Simultaneously, the 2D peak intensity, $I(2D)$, became strongly suppressed, with $I(2D)/I(G)$ ratio changing from ~0.8 before intercalation to ~0.5 after. This behaviour indicates strong electron doping [32-34]. Both peaks returned to the initial position and intensity after deintercalation. The spectra in Fig. 5 were collected during the 2$^{nd}$ intercalation cycle, and similar curves obtained for the 3rd cycle, with the Li ion density reaching stage I/II (Fig. 3a), or carrier density in the bilayer $n \sim (8-9) \cdot 10^{13}$ cm$^{-2}$. As expected, both G and 2D peaks remain visible for this doping, very similar to the case of stage-II Li intercalation of graphite, where doping was even higher [35]. Although further cycling led to higher electron density, it was not possible to quantify this evolution using Raman spectra because of the visible degradation of the electrolyte under the laser beam and the strongly increased noise.

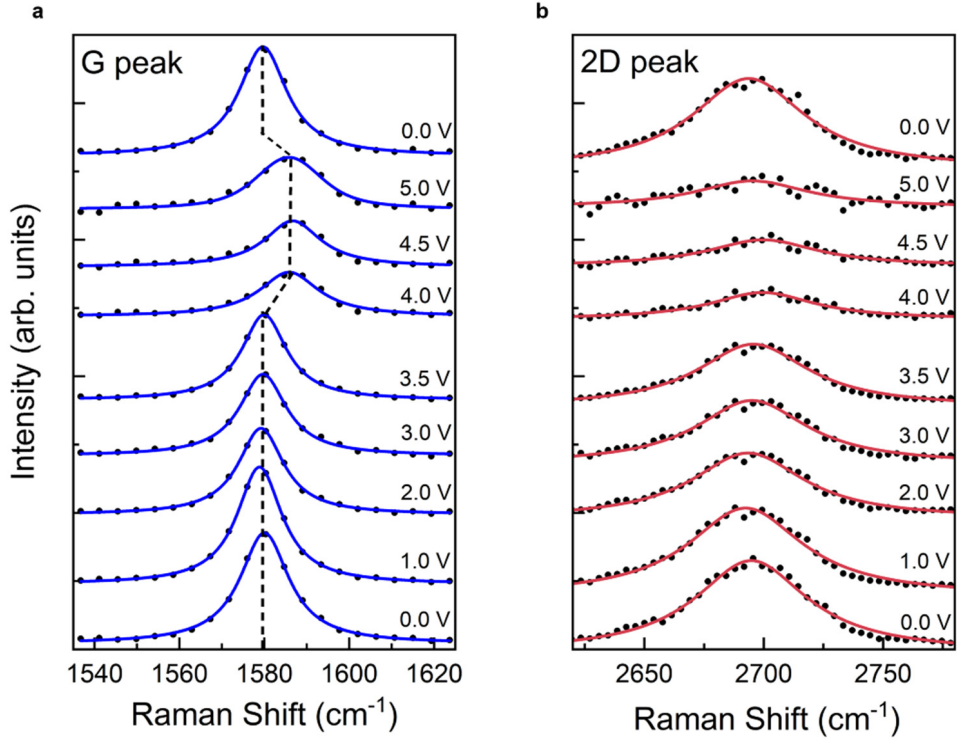

**Figure 5 | Characteristic Raman peaks before, during and after intercalation. (a)** Evolution of the G peak position with intercalation. Individual spectra are shifted for clarity. Corresponding values of the gate voltage, $V_g$, are shown as labels. Intercalated state corresponds to $V_g = 4.0$, 4.5 and 5.0 V. Black dashed line emphasizes the shift of the G peak in the intercalated state. **(b)** Same as in (a) for the 2D peak.

The Raman spectra provided a new insight into the intercalation process: Firstly, no D peak above the noise level was observed either before, during or after intercalation, suggesting that intercalation did not induce point defects in our BLG, in agreement with our conclusions from resistance measurements (no increase in resistance after deintercalation). It is possible that the D peak was comparable in intensity with the noise level, which would give $I_D/I_G < 0.02$ and a minute possible density of point-like defects of $< 5 \cdot 10^9$ cm$^{-2}$ (Supplementary Methods 1.4). Secondly, from the observed shift in the G peak position and the change in $I(2D)/I(G)$ intensity ratio we can place a lower limit on the electron doping induced by intercalation in our devices. As they were always significantly p-doped in the unintercalated state ($n \sim -10^{13}$ cm$^{-2}$), the shift in the G peak position by ~8 cm$^{-1}$ corresponds to electron doping much larger than $10^{13}$ cm$^{-2}$ and larger than observed for electrostatically gated bilayer graphene [33]. Notably, the carrier density $n \sim 5 \cdot 10^{13}$ cm$^{-2}$ achieved in ref. [33] is the highest reported in the literature for the experimental evolution of the G and 2D peaks in BLG but twice lower than for our in-plane stage I/II where the Raman spectra were taken. The data scatter in ref. [33] and the saturation in Raman peaks positions at high electron doping do not allow extrapolation to higher densities in order to estimate the doping level in our devices. That said, we note that both the G peak shifts and carrier density $n$ extracted from the Hall resistance in our experiment agree with those reported for Li ion intercalation of BLG in a similar setup in ref. [9].

For one of the devices (Supplementary Fig. 7), we estimated the amount of Li entering during intercalation from a step in the cathodic current between the counter electrode and the bilayer, corresponding to a step in $\rho_{xx}$ (Supplementary Fig. 7b). Integrating the current step gives a charge transfer $\Delta Q \approx 4 \cdot 10^{-10}$ C, i.e., entry of $\Delta Q / e \cdot S \approx 5 \cdot 10^{14}$ cm$^{-2}$ electrons, in good order-of-magnitude agreement with the number of intercalated Li ions corresponding to the transition from stage I to stage III in this case (~$1.3 \cdot 10^{14}$ cm$^{-2}$, Fig. 3a). Here $e = 1.6 \cdot 10^{-19}$ C is the electron charge and $S \approx 5 \cdot 10^{-6}$ cm$^2$ the area of our ~10μm x 50 μm device.

## Discussion

The distinct in-plane staging observed in our experiments seems to disagree with the earlier reports for BLG [5,6,9] and with the behavior known for Li intercalation of bulk graphite. In the latter case, it has been shown that, after first Li islands form between graphene planes, the resulting local strain leads to attraction between neighboring islands, which promotes further intercalation, so that the full capacity of an interlayer gallery is reached as soon as it is opened by first intercalating islands [28] or the system phase-separates into a 3D checkerboard structure [36]. As for the existing theory, computational studies usually start with assuming a change in stacking of the constituent graphene layers from the AB to AA configuration, which occurs concurrently with Li-ion intercalation [17,22,27]. Such a single structural transition cannot explain the occurrence of the observed four stages.

The only plausible explanation for the four well-defined plateaus in $\rho_{xx}$ and $n_{Li}$ separated by sharp transitions is distinct configurations into which Li ions rearrange themselves at different stages. The constant $\rho_{xx}$ also means that only one phase is present at each plateau: If domains of two different phases were present, e.g., an unintercalated and intercalated one, or domains with different Li ion densities as in graphite [36], we would observe a smooth evolution of $\rho_{xx}$ as the higher density domains grow in size [37]. Distinct ion arrangements in our BLG-Li system are perhaps not surprising because this happens for graphite intercalation where Li ions reside at centers of next-nearest carbon hexagons, creating a hexagonal superlattice. Accordingly, some (at least, short-range) order can also be expected for the BLG. Indeed, Li ions should not only occupy their energetically most favorable sites at carbon hexagon's centers but also would tend to be spaced equidistantly, to minimize the electrostatic energy due to Coulomb repulsion [38]. Importantly, Coulomb repulsion between Li ions is greatly enhanced in BLG with respect to bulk graphite because of much reduced screening in 2D [39,40]. This should favor Li-ion ordering at longer distances [41]. The suggested commensurability between Li ions and the carbon lattice (rather than random Li positions) is consistent with the fact that disorder played little role in our experiments. Indeed, the reproducibility of $\rho_{xx}$ within a few % over many intercalation cycles and for different devices indicates that the observed $\rho_{xx}$ values are intrinsic, being determined by electron-phonon scattering, rather than by disorder induced by Li doping. This conclusion about dominant electron-phonon scattering for our Li-intercalated BLG (rather than other scattering mechanisms) agrees with the previous report for heavily-doped monolayer graphene which showed that its resistivity at room temperature was phonon-limited [42]. The monolayers exhibited $\rho_{xx}$ of ~100 Ohm at similar doping (~$10^{14}$ cm$^{-2}$), which is a factor of 2 lower than $\rho_{xx}$ in our experiments. However, electron-phonon scattering in BLG should be stronger than in monolayer graphene because of a contribution from shear phonons representing local lateral displacements of the two layers [44] leading to a larger phonon-limited $\rho_{xx}$ compared to the monolayer, as indeed observed for our intercalated BLG.

Assuming commensurability between Li ions and the graphene lattice, the measured values of $n_{Li}$ yield the following stoichiometric compositions: $C_{42}LiC_{42}$, $C_{38}LiC_{38}$, $C_{18}LiC_{18}$ and $C_{14}LiC_{14}$ for stages I to IV, respectively (Fig. 3a), where the subscript in $C_NLiC_N$ corresponds to the number $N$ of carbon atoms in each of the two graphene layers per Li ion (see Supplementary Methods 1.3 for details on how the compositions were assigned). Importantly, the inferred stoichiometries correspond to equidistant positions of Li ions at carbon hexagons' centers, which minimizes the Coulomb interaction energy (see Fig. 3b). Our experimental accuracy in determining $n_{Li}$ – which can be judged from the data scatter in Fig. 3a – is sufficient to unambiguously rule out stoichiometries any other than $N$ = 18 and 14 for stages III and IV, respectively. Indeed, the nearest commensurable configurations ($N$ = 8 and 24) correspond to greatly different $n_{\text{Li}}$. Even considering structures with broken hexagonal symmetry in Li-ion arrangements ($N$ = 12, 16 and 20) (hence, unequal separations between Li ions and a larger interaction energy) would result in $n_{\text{Li}}$ well beyond our experimental error ($n_{\text{Li}}$ = 3.2, 2.4 and 1.8×10$^{14}$, respectively). For less-doped stages I and II with the inferred $N$ = 42 and 38, the nearest alternative commensurable configurations are $N$ = 32 and 50, which would result in clearly different doping. However, in the latter case, we cannot fully exclude Li ion arrangements with broken hexagonal symmetries (e.g., $N$ = 36).

To gain further insight into the observed in-plane staging, we used DFT calculations of the Gibbs free energy $\Delta G$ (relative to the unintercalated state) that can be written as (for details, see Supplementary Note 2.1)

$$\Delta G = \rho E_{\text{int}} - \rho \Delta \mu + k_B T \left[ \bar{\rho} \ln(\bar{\rho}) + (1 - \bar{\rho}) \ln(1 - \bar{\rho}) \right], \qquad (1)$$

where $E_{\text{int}}$ is the intercalation energy per Li ion, $\rho = N_{\text{Li}}/N_{\text{C}}$ with $N_{\text{Li}}$ and $N_{\text{C}}$ being the number of Li and carbon atoms in the BLG supercell, and $\bar{\rho} = \rho N_{\text{C}}/N_{\text{sites}}$ (here $N_{\text{sites}}$ is the density of lattice sites available for Li ion intercalation). Equation 1 includes not only changes in the internal energy during intercalation of BLG but also the configurational entropy for ion insertion, where the third term takes into account the difference in the available intercalation sites, $N_{\text{sites}}$, for AB and AA stacking. This term becomes important at low doping levels (Supplementary Note 2.1 and Supplementary Fig. 8a). The second term in Eq. 1 arises because of the difference $\Delta \mu$ in chemical potentials of Li ions in the source (electrolyte) and within the intercalated BLG. The value of $\Delta \mu$ can be determined experimentally as the drop in the pseudo-reference potential $V_{\text{ref}}$ (Fig. 1a) from the start of Li ions' entry into the bilayer to full intercalation. We found $\Delta \mu = 0.4 \pm 0.02$ eV (Supplementary Note 2.1 and Supplementary Fig. 7a). The contribution from the second term increases proportionally to $n_{\text{Li}}$ as per Eq. 1 and is most important for stages III and IV (Supplementary Fig. 8b-d).

The results of our calculations for $\Delta G$ are summarized in Fig. 6 where we focus on Li-ion configurations with high symmetry and, especially, the hexagonal ones that provide equally spaced Li ions and, therefore, lowest contributions to the Coulomb energy (Supplementary Note 2.1). Equidistant Li configurations, shown as bright symbols in Fig. 6, provide local minima in $\Delta G$ with respect to a multitude of other possible more disordered configurations, even if Li ions are allowed to reside only in carbon hexagons' centers ($\Delta G$ for some of the latter configurations are shown as semi-transparent symbols). Furthermore, Fig. 6 shows that the global minimum in the Gibbs energy, which represents the thermodynamic equilibrium for intercalated BLG, occurs at $N$ = 14 ($n_{\text{Li}} \approx 2.7 \times 10^{14}$ cm$^{-2}$), in excellent agreement with the experimentally observed stoichiometry for in-plane stage IV. Note that this Li ion density (storage capacity) is considerably (2.4 times) lower than that reached for stage-1 intercalation of bulk graphite [18-20]. An equally densely packed configuration for BLG, $C_6LiC_6$, is energetically unfavorable, presenting a very significant energy loss. Qualitatively, the lower saturated Li density achievable in BLG can be understood as the result of weak screening of interionic Coulomb repulsion [41] by the two graphene sheets, which makes it energetically costly for Li ions in BLG to reside as close to each other as in graphite. For confirmation, we calculated the intercalation energy for stage-1 graphite and $C_6LiC_6$ composition in BLG (Supplementary Note 2.1). This yielded a considerable difference of ~20 meV per Li ion between intercalation energies for bulk graphite and BLG, respectively. A more extended discussion of the effect of screening can be found in Supplementary Note 2.4.

According to the energy diagram in Fig. 6, AB stacking remains energetically favorable only at low doping ($n_{\text{Li}} < 0.7 \times 10^{14}$ cm$^{-2}$) whereas the broad energy minimum for this stacking occurs at somewhat higher Li densities, between ~0.7 and 0.9×10$^{14}$ cm$^{-2}$. Although AB stacking seems energetically unfavorable with respect to AA stacking for doping levels where the in-plane staging was observed ($n_{\text{Li}} \geq 0.9 \times 10^{14}$ cm$^{-2}$), note that no gradual transition between AB and AA is expected during intercalation because the transition requires nucleation of AA domains within the AB bilayer, which involves local strain. To create such an AA domain in AB-stacked BLG, the gain in intercalation energy $\Delta E(r) = -\sigma A(r)$ should exceed the energy penalty associated with elastic deformations along the domain perimeter, $\gamma C(r)$, where $r$ is the radius of a domain, and $A$ and $C$ are its area and circumference, respectively. The DFT results yield $\sigma(n_{Li}) \approx 2.6$ meV/Å$^2$, and $\gamma$ was estimated as ~ 0.16 eV/Å (Supplementary Note 2.2). This corresponds to the critical domain size $r_c = 2\gamma/\sigma \approx 12$ nm, see Supplementary Fig. 9. The energy barrier for creating such critical-size precursors should postpone the transition from AB into AA stacking and allow the observation of metastable states at $n_{Li} > 0.7 \times 10^{14}$ cm$^{-2}$ and until Li density reaches the value ~1.0×10$^{14}$ cm$^{-2}$ as indicated by the arrow in Fig. 6. Therefore, we attribute in-plane stages I and II to the broad energy minimum seen in Fig. 6 and Supplementary Fig. 8b-d for the AB stacking (black curves). This corresponds to stoichiometric doping with $N$ = 40 ±2 and agrees well with the experimentally found $N$ (Fig. 3a). The above results also allow us to understand the step-like changes in Li concentration from ~10$^{14}$ to > 2×10$^{14}$ cm$^{-2}$ as the structural transition from AB to AA stacking, which is also accompanied by relaxation of Li ions into energetically preferable configurations commensurate with the underlying graphene lattice. In Fig. 6, this transition corresponds to the move between the main minima on the black and blue curves.

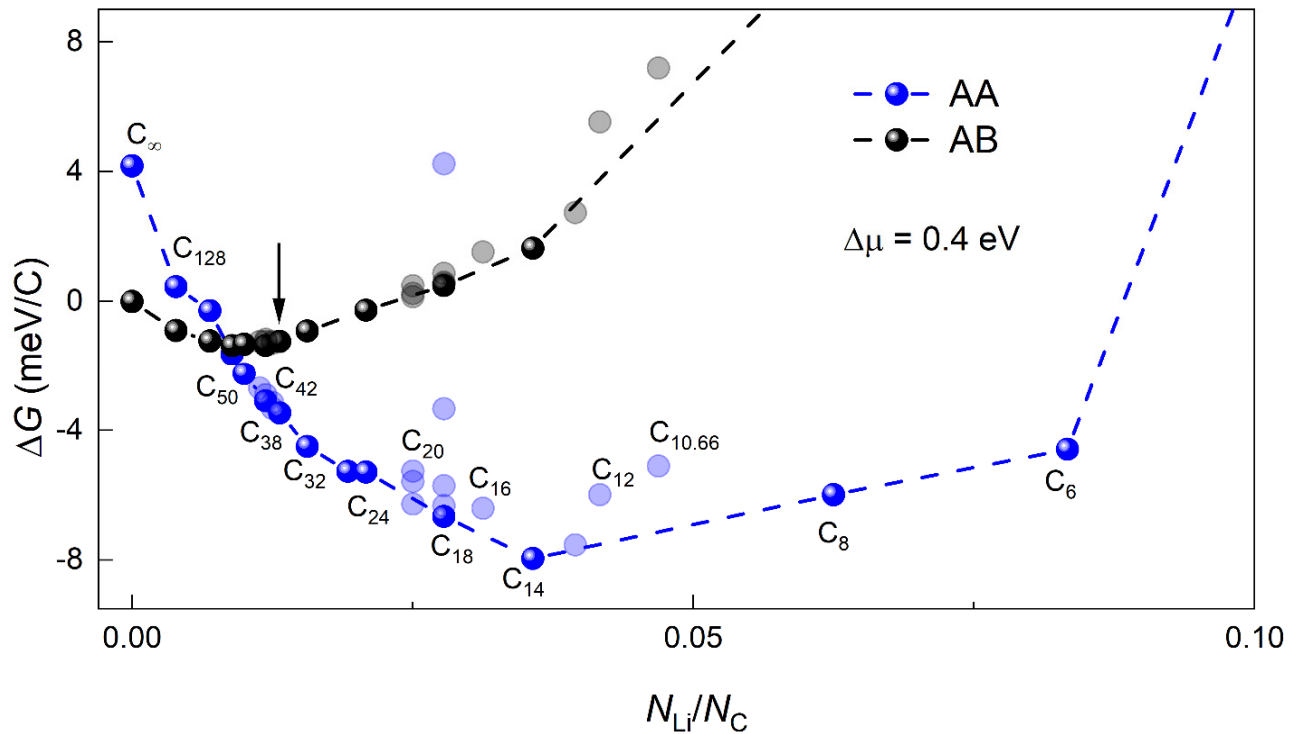

**Figure 6 | Calculated free energy for AB- and AA- stacked bilayer graphene intercalated with lithium.** Data for AB- and AA- stacking are shown by black and blue symbols, respectively. Temperature 330 K and $\Delta\mu = 0.4$ eV, as determined experimentally. Labels mark $N$ in the corresponding $C_N$LiC$_N$ stoichiometries. Bright symbols correspond to Li-ion configurations with the hexagonal symmetry; light symbols to non-hexagonal configurations with non-equidistant Li positions such as, e.g., $\sqrt{21} \times 4$ for C$_{40}$LiC$_{40}$, $\sqrt{13} \times \sqrt{7}$ for C$_{20}$LiC$_{20}$. Where several possible configurations were considered for the same stoichiometry (e.g., C$_{42}$LiC$_{42}$, C$_{20}$LiC$_{20}$, C$_{18}$LiC$_{18}$) the hexagonal ion arrangement was always found to have the lowest energy. Dashed lines: guides to the eye connecting data for the hexagonal arrangements. The arrow indicates $N_{Li}/N_C$ at which the transition from AB to AA stacking is calculated to occur.

The origin of the staging around these two main equilibria (that is, the differences between stages I and II and stages III and IV) is less clear. Our DFT calculations in Fig. 6 yield that stages I and II (attributed to commensurate Li configurations with $N$ = 42 and 38, respectively) have close energies and, therefore, it is reasonable to expect that these energy minima can be occupied depending on fine details of the intercalation process. Because stage I is observed mostly during early intercalation cycles whereas stage II occurs after multiple cycles (Fig. 3a), we speculate that the elastic strain induced during intercalation-deintercalation leads to a slight shift between the two local energy minima around the global one for AB stacking. The strain appears because of the inevitable formation of AB and BA domains during restacking of the graphene layers from AA configuration back into AB configuration during deintercalation (as previously visualized for the case of thermal cycling of BLG [44]), see schematic in the inset of Fig. 4. Furthermore, AB/BA boundaries and, particularly, their intersections containing AA-stacked areas [44-46] can serve as nuclei for restacking AB bilayers into AA configuration, which are needed for reaching intercalation stages III and IV. The development of the extensive network of AB/BA boundaries can also contribute to the 'initial training' effect observed in our experiments in the first few intercalation cycles as well as an analogous behavior known for graphite-based electrodes [47,48]. Indeed, AB/BA boundaries exhibit a larger spacing between the two graphene layers [16], which reduces interlayer adhesion and the energy barrier for Li entry and diffusion [21,25], thus allowing the system to quicker reach stable configurations in subsequent cycles. This can explain why, after several cycles, the time spent during intercalation in AB stacking becomes shorter (Fig. 2c). It is also instructive to note that AB/BA boundaries contain 1D electronic channels with a finite carrier density even at the graphene's neutrality point [45]. Accordingly, this metallic network developed after multiple cycles should provide a notable electrical conductance within the poorly conducting deintercalated state. This additional conductance explains the counterintuitive observation discussed above that $\rho_{xx}$ in the fully deintercalated state decreased after each cycle.

As for stages III and IV that occur within the AA stacking, the DFT calculations yield that $N$ = 18 and 14 are also rather close in energy (Fig. 6). Because stage III was often found gradually transforming into stage IV (Supplementary Fig. 3c), we suggest that stage III ($N$ = 18) corresponds to a long-lived metastable state. The slow transition between stages III and IV also agrees with the fact that Li ions experience radically different diffusion barriers for AB and AA stacking, which are calculated as ~70 and 280 meV, respectively (Supplementary Fig. 10). These values suggest rapid diffusion and, hence, sharp transitions between stages I and II whereas exponentially longer times can be required to reach local equilibria in the case of AA stacking (that is, to move between stages III and IV).

We emphasise that an important difference between intercalation of graphite and of our small, defect-free graphene bilayer devices is the timescale. In graphite intercalation typically takes hours, while in BLG Li ions fill the whole device in one step due to ultrafast diffusion [5]. In principle, one could expect the bilayer to go through all the configurations identified in our DFT calculations as energy minima (Fig. 6), but we only see the most stable ones, with 'jumps' between them due to significant Li density differences and ultrafast diffusion. The 'jumps' happen at a constant gate voltage due to finite energy barriers between different $C_xLiC_x$ configurations: the largest barrier is for AA domain formation and restacking from AB to AA and an appreciable barrier for Li diffusion through the AA-stacked bilayer. The time span of each $\rho_{xx}$ plateau then depends on the value of overpotential (Supplementary Fig. 4) and with repeated cycling is also affected by developing non-uniformities as discussed above. Abrupt jumps only occurred between stage I/II (AB stacked bilayer) and stage III/IV (AA stacked), supporting this explanation. The transition from stage III to stage IV was typically smooth (Supplementary Fig. 3c), while a transition from stage I to stage II was observed only once (Fig. 3a).

## Conclusions

Our work was initially motivated by the lack of understanding of what determines the limits for Li intercalation in bilayer graphene. Although the answer may seem disappointing for potential applications, it is important that future developments take into account that the superior conductivity, large surface area, and ultrafast Li diffusion in potential ultrathin graphene electrodes would be tempered by a reduced Li storage capacity. This is particularly relevant for dense assemblies of BLG considered for battery technologies, which could provide larger storage capacity than the one observed for isolated bilayers.

On a more fundamental level, we have identified previously unknown essential characteristics of the intercalation process. We have demonstrated that intercalation occurs in AB-stacked bilayers without immediate restacking to the AA configuration; AA restacking requires achieving a finite, rather large, Li ion density and is itself required to achieve saturation in Li content. The two stages for each stacking order (AA and AB) involve relatively small changes in Li concentrations and are attributed to local equilibria that are close in energy and occur either as metastable states or because of shifting equilibrium conditions during intercalation cycles. We find that BLG can provide only weaker screening of interionic interactions compared to bulk graphite, so that Li ions interact stronger and start repelling each other at longer distances, limiting the storage capacity of BLG. Another surprising finding is the experimental evidence for highly ordered Li configurations (essentially Li ion superlattices) which is of interest for electronic transport properties. It would be interesting to visualize the suggested $C_xLiC_x$ configurations by other techniques, especially scanning tunnelling microscopy.

**Data availability.** The authors declare that the data supporting the findings of this study are available within the paper and its supplementary information file.

# Supplementary Information

## 1. Supplementary Methods.

### 1.1. Device fabrication

To fabricate BLG devices, such as shown in Figure 1a of the main text, a bilayer graphene crystal was mechanically exfoliated from bulk graphite and transferred onto a Si/SiO$_2$ (290nm) substrate. This was followed by deposition of metal contacts using standard electron-beam lithography with poly(methyl methacrylate) (PMMA) resist and e-beam evaporation of Cr (3nm)/Au (40nm). In the next step a similar lift-off procedure was used to fabricate a large counter electrode made of Ta (3nm)/Pt (40 nm) at ~ 250 μm distance from the BLG and two reference electrodes, also Pt, see Supplementary Fig. 1a. Finally, the bilayer was shaped into a Hall bar geometry using a PMMA etch mask and reactive ion etching in oxygen plasma. To ensure that only a small part of the bilayer is exposed to the electrolyte during intercalation, we used a protective layer of SU-8 3005 (~ 5 μm thick) as shown in Fig. 1b and Supplementary Fig. 1a. Owing to the excellent chemical stability and insulating properties of SU-8 [1], the device and the Cr/Au contacts were protected from unwanted electrochemical reactions that could take place due to contact with the electrolyte. The design of the SU-8 layer also ensured that Li ions from the electrolyte could enter the bilayer only through the exposed edge of the device.

### 1.2. Electrolyte preparation and characterization

We used solid polymer-based Li-ion electrolyte PEO-LiTFSI [2,3]. The electrolyte was prepared in an Ar-filled glovebox. Prior to mixing the ingredients, a lithium salt, bis(trifluoromethanesulfonyl)imide (LiTFSI), and poly(ethylene oxide) (PEO) (M$_W$ = 100,000 g/mol) were dried overnight at 170°C and 55°C, respectively, and acetonitrile was dehydrated with 3 Å molecular sieves at room temperature. Then 0.10 g LiTFSI and 0.31 g PEO were mixed with 2 mL acetonitrile under continuous stirring at room temperature for at least 24 hours, giving the molar ratio of ethylene oxide to lithium (EO/Li) of 20:1 and a viscosity suitable for drop-casting. To this end, 2 μL of the pre-mixed electrolyte was drop-cast over the device using a micropipette, ensuring that the electrolyte covered the area encompassing the Pt counter and pseudo-reference electrodes, the exposed part of the bilayer graphene and a part of the SU-8 protective layer, but none of the exposed Au contact pads. The electrolyte was solidified through evaporation of acetonitrile at 50°C overnight to form Li ion conductive solid polymer electrolyte. The ionic conductivity of this electrolyte is due to Li ions moving under the applied voltage, as they migrate between oxygen cites on the PEO backbone [3]. The prepared electrolyte was characterized using standard impedance spectroscopy [3] in a Pt–electrolyte–Pt two-probe configuration, in a frequency range 10 Hz – 500 kHz. Typical Nyquist plots of the electrolyte impedance at several different temperatures are shown in Supplementary Fig. 1c,d. Additionally, we used impedance spectroscopy to determine the internal resistance of our devices, which was found to be $R_0 \approx 20$ kΩ (see Supplementary Fig. 1b for an example). To analyze the measured spectra, the electrochemical system was modelled as an equivalent electrical circuit shown in Supplementary Fig. 1b, using ZView® software. Measurements at temperatures between 30°C and 60°C showed that the ionic conductivity of our electrolyte is strongly temperature-dependent, increasing 50 times as temperature increased from 30°C and 56°C. Increasing the temperature further, from 56°C to 60°C, had a much smaller effect. Taking into account the melting point of the electrolyte, $T_\mathrm{m} = 65$°C, we have chosen 57°C as optimal for our intercalation/deintercalation measurements.

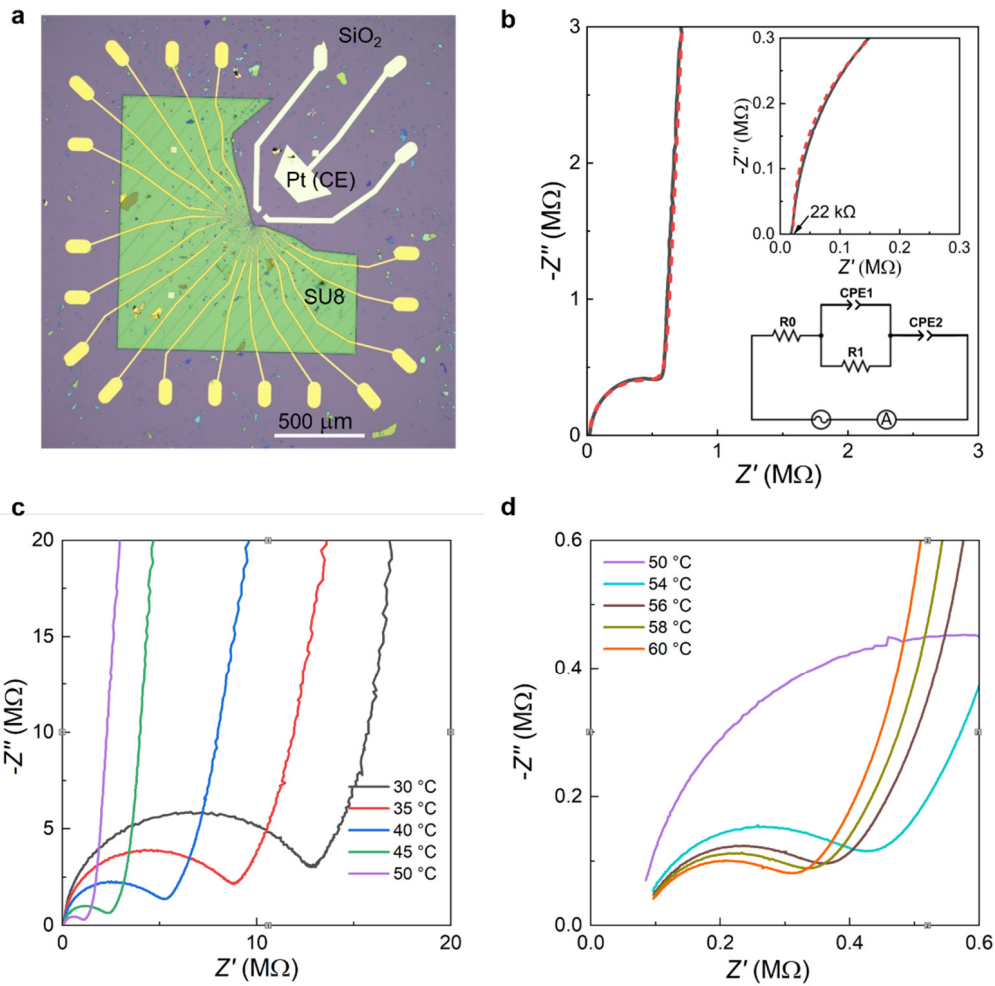

**Supplementary Figure 1 | Device overview and impedance spectroscopy characterization. (a)** Optical image of the device assembly showing Pt counter- and pseudo-reference electrodes (white) and the protective SU8 layer (green). **(b)** Nyquist plot of the device impedance measured between the Pt counter electrode and BLG at $T =57$ °C (black solid line) and the equivalent circuit fit obtained using ZView® software (red dashed line). The inset shows a zoom of the high-frequency part indicating the internal resistance of the device, $R_0 \approx 22$ kΩ. Bottom right inset: equivalent circuit. **(c,d)** Nyquist plots of the electrolyte impedance measured between the counter- and reference Pt electrodes at different temperatures, see legends.

### 1.3. Intercalation and transport measurements

A schematic of the on-chip electrochemical cell used in our experiments is shown in Fig. 1a (main text) and the image of a finished device in Supplementary Fig. 1a. To achieve intercalation, the gate voltage $V_g$ was swept from 0 to -7V (or a lower value, as described in the main text) at a constant rate of 10 mV·s$^{-1}$. For deintercalation, $V_g$ was returned to 0V at the same rate. We have used Pt both for the counter- and the pseudo-reference electrodes; the source of Li ions in our electrochemical cell is the LiTFSI salt in the electrolyte. The role of the counter electrode is to create a potential drop at the interface with BLG, such that the chemical potential of Li in $C_x LiC_x$ becomes lower than the chemical potential of 'free' lithium in the electrolyte (with an unknown chemical potential). The threshold potential for intercalation in this case does not correspond to any thermodynamic value (as in the case of Li/Li$^+$ electrode) and was determined experimentally for each device. To compare the experimental results with our DFT calculations of the free energy, we defined a reference potential as $\mu_{\text{ref}} = eV_{\text{ref}} = \mu + \Delta\mu$, where $V_{\text{ref}}$ is the measured pseudo-potential and $\Delta\mu = 0$ corresponds to the start of intercalation (sharp peak in the device resistance, see below). This allowed us to determine $\mu$ experimentally as $\mu \approx eV_{\text{ref}}$ at $\Delta\mu = 0$. For example, for a device in Supplementary Fig. 7a below we found $V_{\text{ref}} \approx 2.9$ V, that is, $\mu_{\text{ref}} = \mu \approx 2.9$ eV. The applied $V_g$ in this case was -4.5 V; the difference between $V_g$ and $V_{\text{ref}}$ is due to a voltage drop at the counter electrode.

We note that, although $V_g = -7$ V nominally exceeds the electrochemical stability window for PEO-LiTFSI electrolyte [4], we found that a large fraction (30 to 40%) of the applied 7V potential difference falls at the Pt counter electrode, where no faradaic reactions are expected. The potential drop at the graphene interface (as found from our measurements of the potential difference between the Pt pseudo-reference and graphene) is then < 4.5V, comparable or less than the reported stability window ~4.5 V vs Li/Li[+] [4]. The additional voltage drop due to the relatively high internal resistance of our devices ($R_0 \approx 20$ kΩ, Supplementary Fig. 1b) was always negligible compared to the applied gate voltage, <0.2 mV, due to very small anodic/cathodic currents in our experiments, $I < 10$ nA. For the number of cycles used in our study (<20) we did not see significant changes in electrolyte conductivity or in the internal device resistance after cycling, as confirmed by repeated impedance spectroscopy measurements. We also did not see any visible evidence of the electrolyte movement when inspecting the devices before and after intercalation.

Due to air sensitivity of the LiTFSI electrolyte and of the intercalated BLG, all experiments were carried out in the inert atmosphere of an Ar filled glovebox. To ensure sufficient ionic conductivity of the electrolyte (see above), the device was kept under constant heating to maintain its temperature at 57°C. The large thickness of the SU-8 layer prevented ionic gating from the surrounding electrolyte that could in principle contribute to changes in the carrier density and the resistance of the bilayer. Control experiments on monolayer graphene in a similar setup confirmed that Li ions did not intercalate between graphene and SU-8; intercalation only occurred in the gallery of the bilayer, in agreement with previous studies.

A magnetic field $B = \pm 330$ mT was provided by a permanent magnet positioned immediately above the device within the glovebox. The density of Li ions, $n_{\text{Li}}$, at each stage of intercalation was found from the carrier density in the BLG, $n$, measured at certain points during intercalation and at the end of each intercalation (deintercalation) half-cycle. Here $n = B \cdot I/(e \cdot V_{\text{Hall}})$ was determined from the Hall voltage measured at $\pm 330$ mT, where $V_{\text{Hall}} = [(V_{\text{xy}}(B) - V_{\text{xy}}(-B)]/2$, $I$ = 1 μA is the applied current and $e$ the electron charge. BLG in all our as-fabricated devices was initially p-doped (carrier density $n_0 \approx$ -1.2 x10[13] cm[-2], where the minus sign is used to indicate hole doping). Such level of p-doping is typical for unencapsulated graphene in contact with polymers [5,6] and, in our case, was probably caused by fabrication residues and/or SU-8 layer [5]. Taking into account the initial p-doping, the density of intercalated Li ions was calculated as $n_{\text{Li}} = (n + |n_0|)/0.9$, where $|n_0|$ is the carrier density in the bilayer at the start of intercalation ($t = 0$) and it is assumed that each Li ion donates 0.9e to the BLG host (see main text). We note that in the literature the DFT charge transfer value for Li in graphite/graphene varies from 0.85 to 0.88 to 0.9e[-]; we are using a rounded value of 0.9e[-] throughout. Using a lower value of 0.88e[-] would result in just 1-2% difference in $n_{\text{Li}}$, which is below our experimental error and did not affect the inferred Li ion configurations or other conclusions. To estimate $n_{\text{Li}}$ for stages II and III, which were too short in duration to measure $V_{\text{xy}}$ at $\pm B$ (open symbols in Fig. 3a), we used the assumption of a constant scattering time $\tau$ in $\rho_{\text{xx}} = m^*/(ne^2\tau)$ for each stage of intercalation, see discussion in the main text ($m^*$ is the electron mass). The carrier density was then estimated as $n \approx A/\rho_{\text{xx}}$, where $A = m^*/e^2\tau$ was assumed to be constant and found as $A \approx \rho_{\text{xx}} n_{\text{meas}}$ for a given stage using both resistance and Hall measurements as described above.

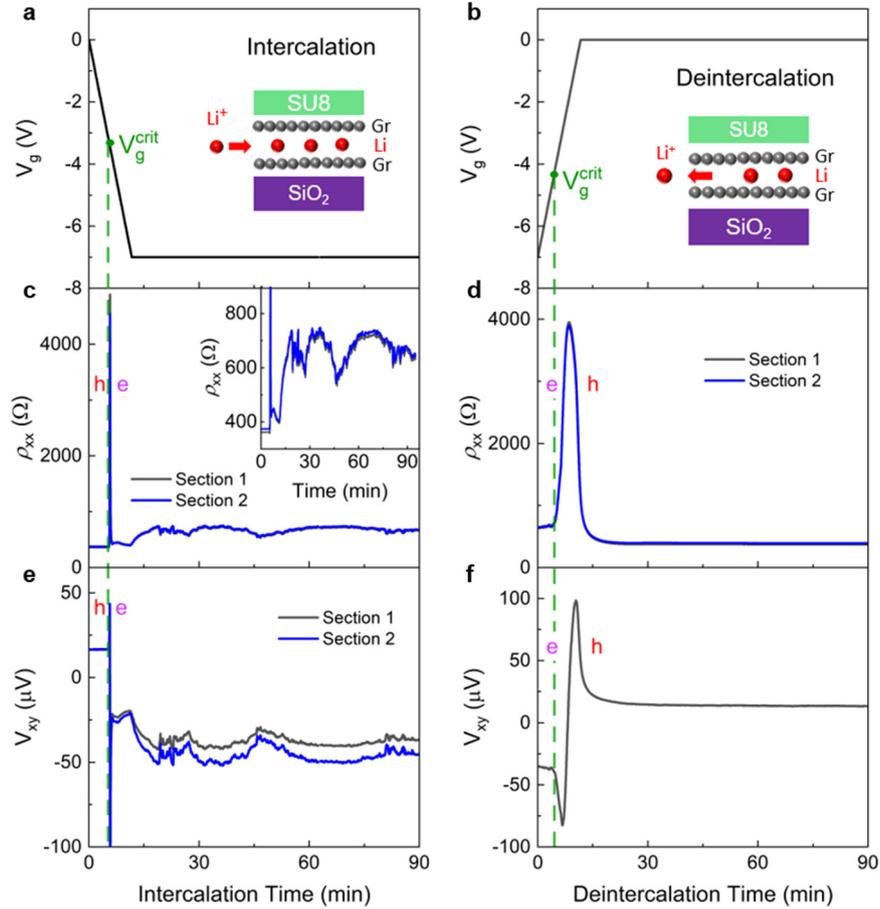

**Supplementary Figure 2 | First intercalation-deintercalation cycle.** Evolution of the longitudinal resistance and Hall voltage of BLG during intercalation (c,e) and deintercalation (d,f). Blue and black curves in (c,e) show $\rho_{xx}$ and $V_{xy}$ measured at two different sections of the device, separated by ~10 μm. Top panels (a,b) show the corresponding sweep of the gate voltage $V_g$ and indicate its critical value corresponding to the start of intercalation (deintercalation).

For completeness, Supplementary Figs. 2&3a show $\rho_{xx}(t)$ and $V_{xy}(t)$ for the 1st intercalation-deintercalation cycle and $\rho_{xx}(t)$ for all 14 intercalation cycles for device A, respectively. In the 1st cycle (Supplementary Fig. 2), as the gate voltage $V_g$ was swept from 0 to $-7$V, $\rho_{xx}$ changed from a steady value of ~350 Ω in the initial p-doped state to ~420 Ω and $n \approx 8 \times 10^{12}$ cm$^{-2}$ in the intercalated state, with the change of graphene's polarity occurring over ~1 min, at $V_g \approx -3$V. This indicated a fast entry of a large number of Li ions corresponding to interionic distances ~2.5 nm, or one Li per 170 carbon atoms. However, after a few minutes $\rho_{xx}$ started to increase and underwent significant fluctuations, even though all external factors remain the same: $V_g$ kept at $-7$ V, temperature at 57°C, etc. Furthermore, the fluctuations that were much larger than the measurement noise (~1Ω) occurred simultaneously across the entire device (cf. $\rho_{xx}$ and $V_{xy}$ for two different pairs of contacts in Supplementary Fig. 2a). In contrast to the intercalation behavior, deintercalation proceeded smoothly and was slower (Supplementary Fig. 2b), with the bilayer returning to its initial p-doped state after approximately 30 min for this first cycle. In the 2nd cycle, $\rho_{xx}$ in the intercalated state remained at a more constant value but fluctuations – and therefore instabilities in the amount of intercalated Li ions – remained. The maximum Li ion density at the end of intercalation in this cycle was $n_{Li} \approx 5.3 \times 10^{13}$ cm$^{-2}$, or one Li ion per ~140 carbon atoms, giving an average composition ~$C_{70}LiC_{70}$, that is, a very dilute and likely disordered intercalation state. It is possible that intercalation in the initial one or two cycles was affected by formation of a solid-electrolyte interphase (SEI) and it may have contributed to the observed changes (in addition to bilayer expansion and formation of AB/BA boundaries discussed in the main text). Importantly, SEI formed at negative potentials is porous and not expected to prevent intercalation [4].

The 3$^{rd}$ cycle was characterized by an almost constant $\rho_{xx} \sim 250$ $\Omega$ maintained over >1h and much smaller resistance fluctuations compared to the 2$^{nd}$ cycle. First signs of stable Li configurations and the development of staging appeared from the 4$^{th}$ cycle onwards as described in the main text.

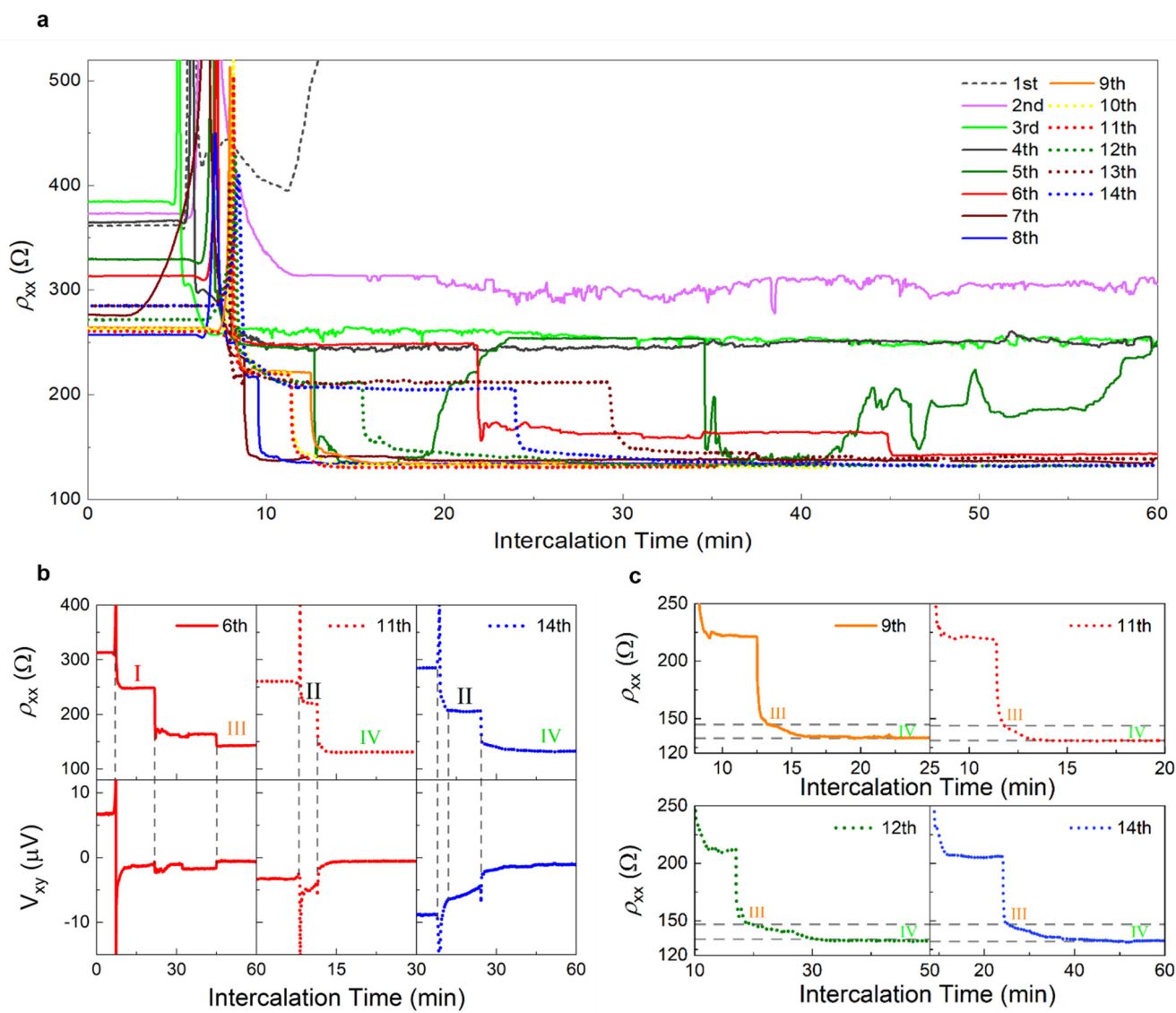

**Supplementary Figure 3 | Intercalation evolution and developing staging. (a)** Evolution of the longitudinal resistance during the first hour in 14 consecutive intercalation cycles (see legends for the cycle numbers). The resistance in the 1$^{st}$ cycle (black dashed curve, not shown after 10 min) remains > 500 $\Omega$ and is shown separately in Supplementary Fig. 1a. **(b)** Simultaneous measurements of the longitudinal resistance $\rho_{xx}$ and Hall voltage $V_{xy}$ during intercalation. The top panels show evolution of $\rho_{xx}$ for the selected cycles and the bottom panels show corresponding $V_{xy}$. The measurements were done under a constant magnetic field of 330 mT. **(c)** Gradual changes in $\rho_{xx}$ signifying the transition from stage III to stage IV (see the main text). Shown are several representative cycles. The dashed lines indicate the $\rho_{xx}$ values for intercalation stages III and IV. All the data are from device A.

Supplementary Fig. 4b demonstrates that achieving the high-density intercalated state (stage IV) did not depend on the magnitude of $V_g$ as long as it exceeded the critical value (typically, $V_g \approx -3.5$ V in our experiment). To this end, in one of the intercalation cycles for device B (Supplementary Fig. 4) we used $V_g = -3.8$ V, well below $V_g = -7$ V used in other measurements. This resulted in a much longer time needed for the transition from stage I to stage IV but did not affect the final result. A later increase of $V_g$ to $-7$ V did not have any effect on the intercalation process. This finding is consistent with the known effect of the overpotential on the interfacial kinetics and Li ion diffusion in graphite (ref. [29] in the main text): at small overpotentials, such as used in this experiment, the limiting factor for Li ion insertion is likely to be the interfacial charge transfer resistance, rather than the diffusion of ions, resulting in slow kinetics.

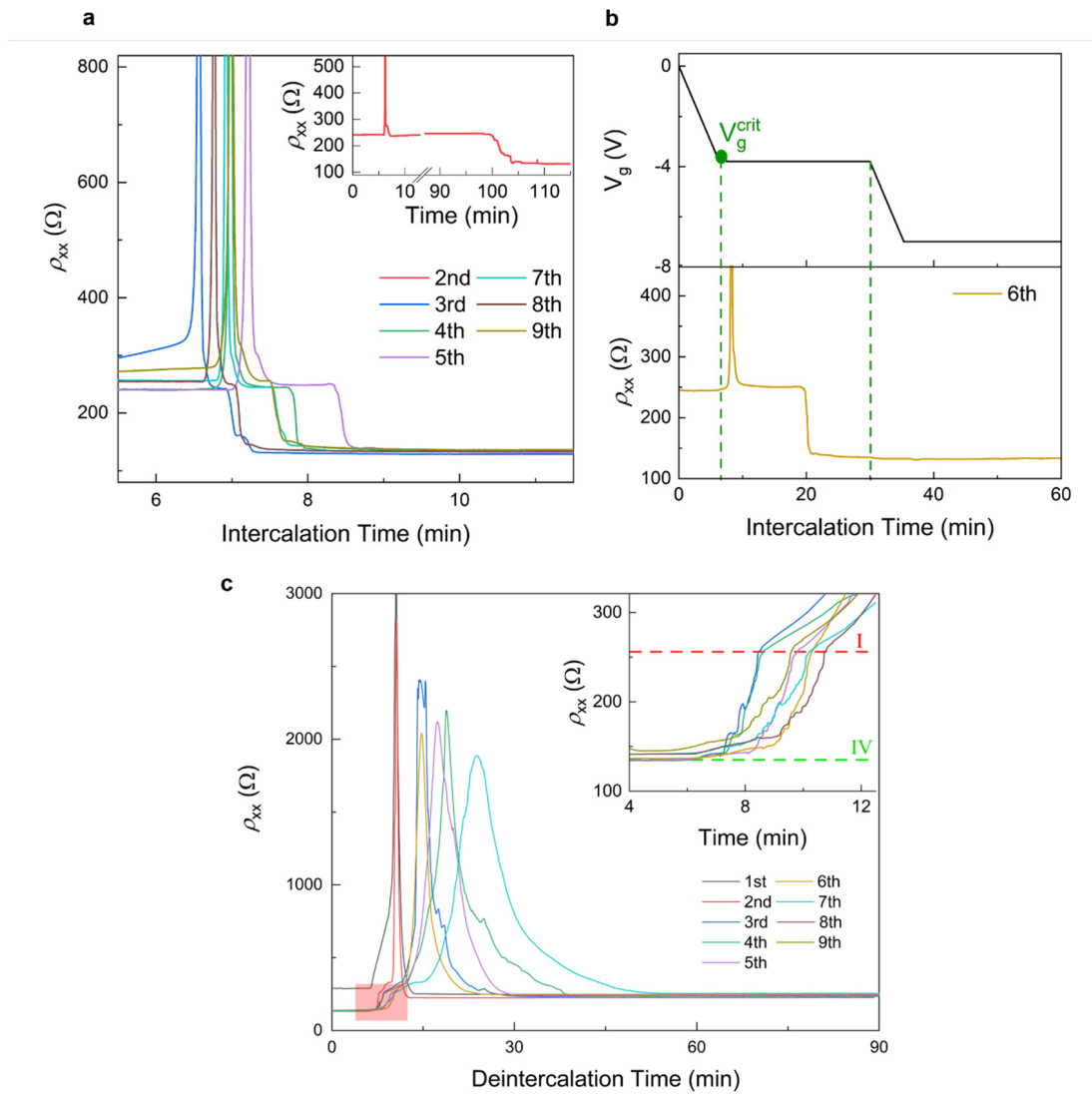

**Supplementary Figure 4 | Further examples of intercalation behavior and the effect of the driving (gate) potential. (a)** Representative intercalation cycles for device B. Longitudinal resistance values in this device are identical (within experimental accuracy) to those for device A (Fig. 1 in the main text), i.e., $\rho_{xx} = 248 \pm 4$ Ω for stage I and $\rho_{xx} = 133 \pm 3$ Ω for stage IV. Also, similar to device A, the transition to a higher stage occurred only after > 100 min in the 2$^{nd}$ cycle (inset). **(b)** Time evolution of $\rho_{xx}$ at a lower driving voltage $V_g = -3.8$ V (just above the threshold value) as opposed to $V_g = -7$ V used in most measurements. The transition from stage I to stage III/IV is seen to occur at $V_g = -3.8$ V but it takes a significantly longer time for this to happen [cf. panel (a)]. No further changes are seen as $V_g$ is increased to $-7$ V. **(c)** Representative deintercalation half-cycles for device B (color coded curves). The inset shows a zoom of the $\rho_{xx}$ curves within the pink square in the main panel. Similar to device A, the transition from stage IV to stage I is clearly seen at the beginning of deintercalation, as well as broadening of the resistance peak and a gradual decrease in the maximum value of $\rho_{xx}$ in later cycles.

For completeness, Supplementary Fig. 5 shows additional data for deintercalation of device A and Supplementary Fig. 6 illustrates the reproducibility of the reported in-plane staging for different devices.

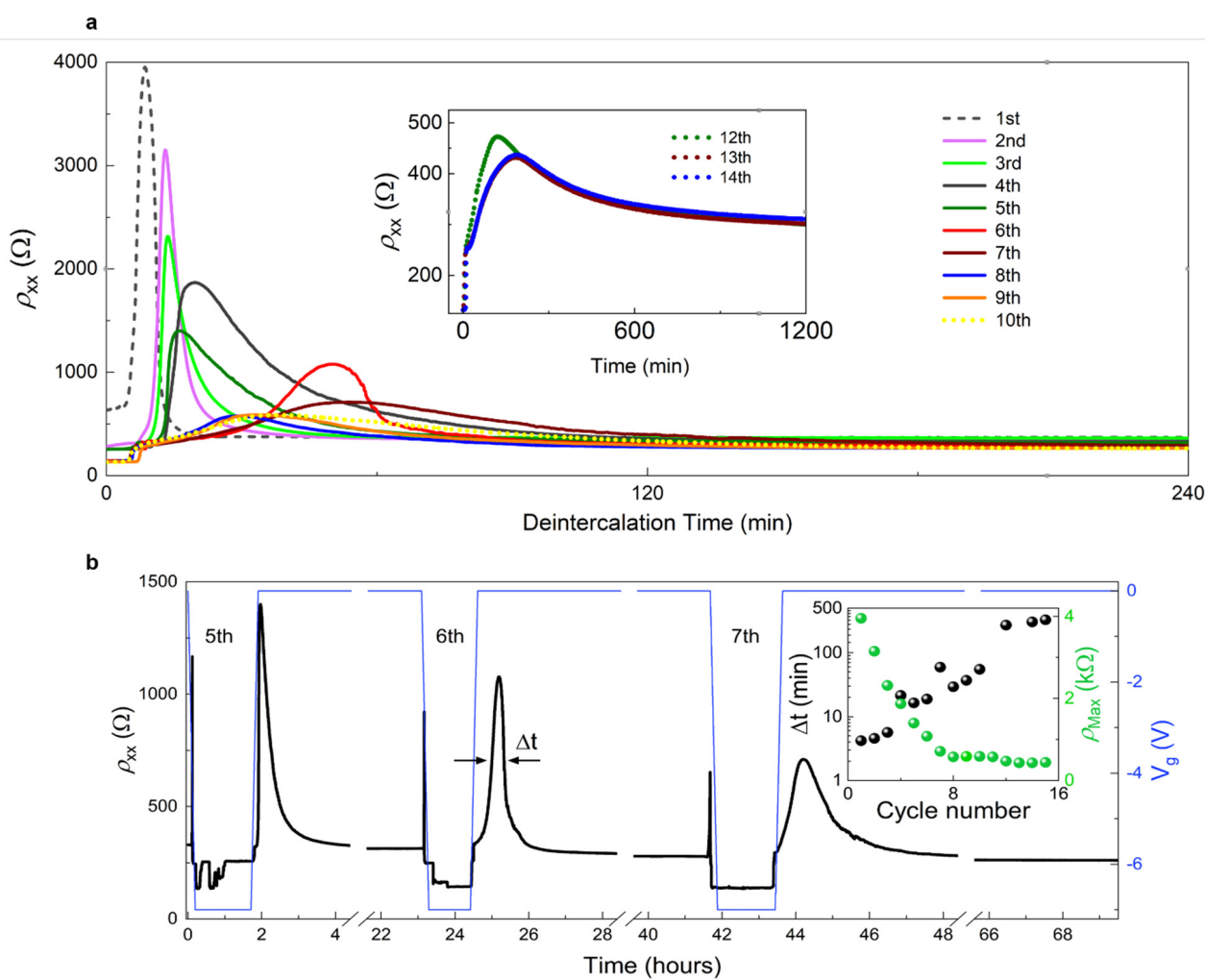

**Supplementary Figure 5 | Evolution of the bilayer resistance during deintercalation. (a)** Time dependence of $\rho_{xx}$ during deintercalation in all consecutive cycles for device A (see legends for cycle numbers). Same color-coding as for the intercalation curves in Figs 1d and S3a. **(b)** Consecutive intercalation-deintercalation cycles (5$^{th}$, 6$^{th}$, 7$^{th}$) illustrating gradually slowing deintercalation dynamics. Blue and black curves show $V_g(t)$ and $\rho_{xx}(t)$, respectively. *Inset:* Changes in the width and height of the deintercalation resistance peak, $\Delta t$ and $\rho_{max}$ respectively, with the cycle number. Data for device A.

To assign stoichiometric compositions to intercalation stages, we calculated Li:C ratios ($N$ in $C_N LiC_N$) using the extracted average Li densities $n_{Li}$: $N = 2/n_{Li}A$, where $A = 5.23 \cdot 10^{-16}$ cm² is the area of a hexagonal unit cell containing 2 C atoms. The values of $N$ calculated in this way for stages I, II, III and IV were 42.0, 37.8, 17.2, 13.9 which were rounded to the nearest integer yielding $C_{42}LiC_{42}$, $C_{38}LiC_{38}$, $C_{18}LiC_{18}$ and $C_{14}LiC_{14}$ (rounding to 18 for stage III ensured Li positions in the centres of C hexagons). Additionally, we have calculated the *expected* $n_{Li}$ for $N$ =42, 38, 18, and 14 as $n_{Li} = 2/N \cdot A$ – these are shown as horizontal lines in Fig. 3a.

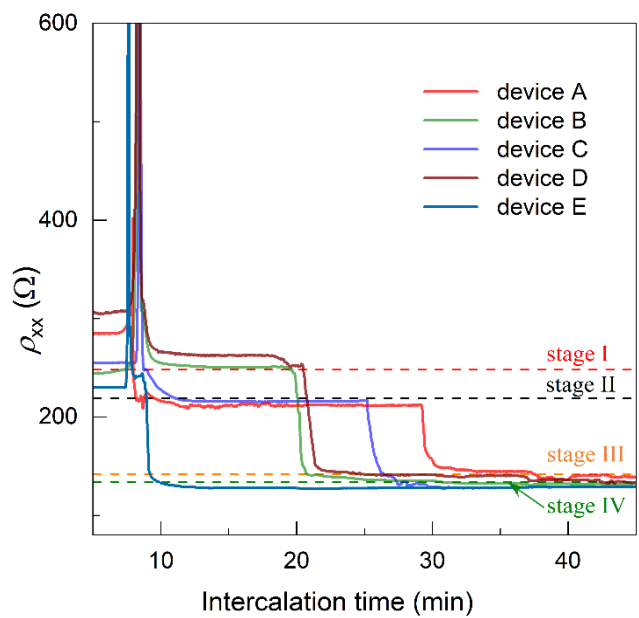

**Supplementary Figure 6 | Reproducibility of staging on different devices.** Time evolution of $\rho_{xx}$ for all 5 studied devices. Shown are representative (middle) cycles, where staging has already been established. As described in the main text, for some of the devices the intermediate plateaus (stages II and/or III) were missing in any of the studied cycles. Additionally, the time duration of the more dilute stages was sometimes very short (e.g., device E), where the bilayer transitioned very quickly from stage I to stage IV.

### 1.4. Raman spectroscopy

As an alternative method to monitor intercalation, we used *in operando* Raman spectroscopy. The device of the same design as in Supplementary Fig. 1 was placed in a sealed custom-made chamber with an optical (quartz) window. A thin layer of the electrolyte was deposited prior to sealing in an Ar-filled glovebox. The gate voltage $V_g$ was swept in small steps while continuously measuring the longitudinal resistance, and a Raman spectrum taken at the end of each step. We used a commercial WITec Raman spectrometer with 514 nm excitation laser, and used three 80s acquisitions per spectrum. Representative spectra before, during and after intercalation are shown in Fig. 5. As noted in the main text, no D peak could be detected either before, during or after intercalation. Nevertheless, it is still possible that a very small D peak (below the noise level) was present, so we estimated the upper limit on the defect density from the $I_D/I_G$ ratio. In deintercalated state, assuming $I_D \leq 30$ (noise level), we obtained $I_D/I_G < 0.02$ and the upper limit on the defect density $n_D < 7.5 \times 10^9 \times E_L^4 \times I_D/I_G \approx 5 \cdot 10^9$ cm$^{-2}$ (here $E_L = 2.4$ eV is the laser excitation energy) [7].

## 2. Supplementary Notes

### 2.1. *Ab-initio* thermodynamic analysis of Li intercalation into BLG

DFT calculations were performed using Quantum ESPRESSO code [8,9]. The Li to C ratio was varied by inserting a single Li atom into BLG supercells of different dimensions, through choice of different crystallographic supercell lattice vectors. A spacing of 20 Å between repeated images in the out-of-plane direction was applied to avoid interactions between these images. GBRV ultrasoft pseudopotentials were used to approximate the effect of core electrons [10], parameterized using the Perdew-Burke-Ernzerhof (PBE) Generalized Gradient Approximation (GGA) of the exchange-correlation functional [9]. The Brillouin zone was sampled using a regular Monkhorst-Pack $k$-point grid of $28 \times 28 \times 1$ in a BLG unit cell [10], and calculations in larger cells were fixed at the same $k$-point density. The optB88-vdW DFT functional was used to model van der Waals interactions [11]. Coulomb interactions in the out-of-plane direction were truncated to avoid spurious long-range interactions. This was found to increase the calculated DFT energies, both for AA and AB stacking of Li intercalated bilayer graphene, see Supplementary Fig. 8a.

To extract the intercalation energy per Li ion from DFT data we used the expression

$$E_{\text{int}} = \frac{(E_{\text{cell}} - N_C \epsilon_{\text{BLG}})}{N_{\text{Li}}} - \mu_{\text{ref}}, \quad (1)$$

where $E_{\text{cell}}$ is the energy of a fully relaxed DFT cell with a given Li density (ratio of the number of Li ions $N_C$ to the number of carbon atoms $N_{\text{Li}}$), $\epsilon_{\text{BLG}}$ is the energy per carbon atom of AB-stacked bilayer graphene, and $\mu_{\text{ref}}$ is a reference chemical potential of intercalated Li ions with respect to 'free' lithium. In previous computational studies, the reference chemical potential ($\mu_{\text{ref}}$) was often chosen with respect to metallic Li [12] or an isolated Li atom [13]. As in our experiment 'free' lithium is in the form of mobile ions in LiFTSI electrolyte (with an unknown chemical potential), we choose $\mu_{\text{ref}} = \mu + \Delta\mu$ and calibrate $\mu$ by calculating the Gibbs free energy such that intercalation of Li ions is energetically disfavored for all Li densities, i.e., such that $\Delta\mu=0$ (corresponding Gibbs free energy as a function of Li density is shown in the inset of Supplementary Fig. 8b).

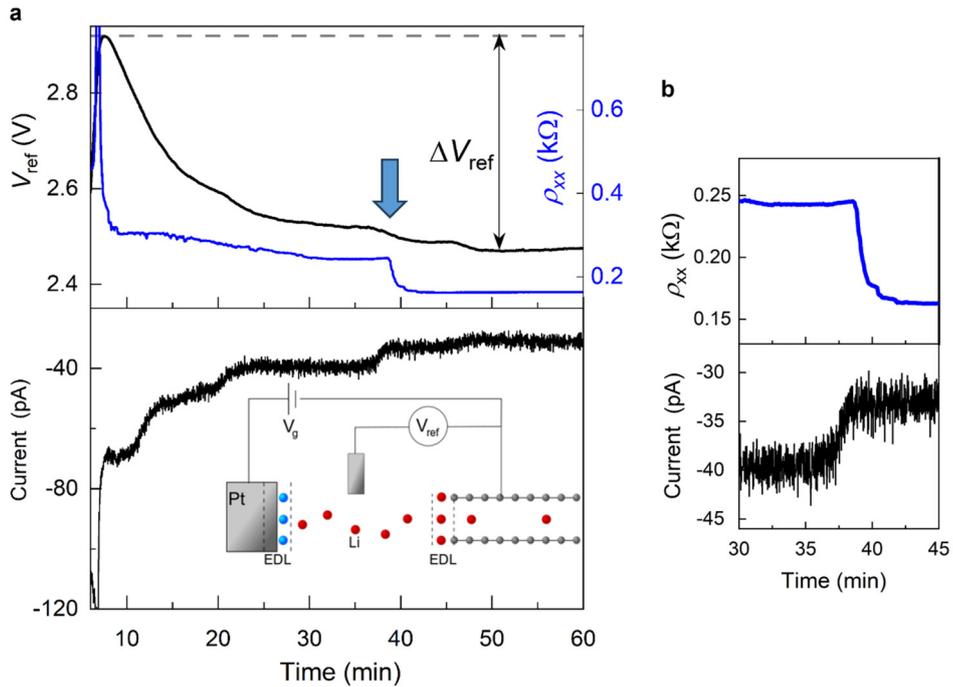

**Supplementary Figure 7 | Experimental determination of the difference in chemical potentials of Li ions in the source (electrolyte) and the intercalated bilayer. (a)** *Top panel*: Simultaneous measurements of the pseudo-reference potential $V_{\text{ref}}$ (black) and the longitudinal resistance $\rho_{xx}$ of the device (blue). The difference in chemical potentials is determined as $\Delta\mu = e\Delta V_{\text{ref}}$, where $\Delta V_{\text{ref}}$ is the drop in reference potential between the start of intercalation (sharp peak in $\rho_{xx}$) and the fully intercalated state, as indicated in the figure. *Bottom panel*: Corresponding current through the electrolyte. *Inset*: Measurements schematic. EDL refers to the electric double layer at the solid-electrolyte interface. **(b)** Zoom of the $\rho_{xx}$ step and the corresponding step in cathodic current used to estimate the charge transfer corresponding to the transition from stage I to stage III (see text).

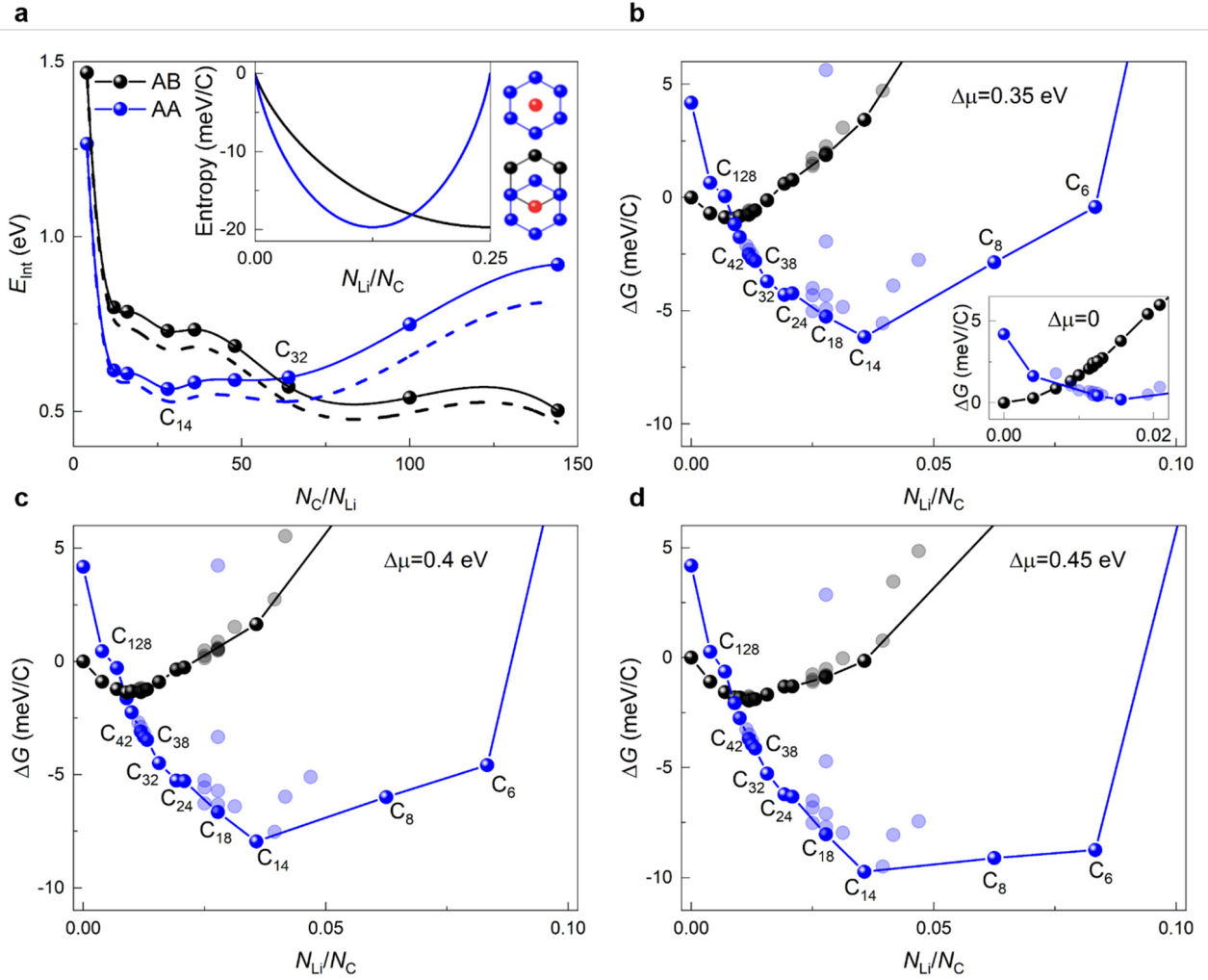

**Supplementary Figure 8 | DFT intercalation energy and Gibbs free energy for reference chemical potentials of Li corresponding to the experiment. (a)** Main panel: Intercalation energy vs the ratio of the number of carbon atoms $N_C$ to the number of lithium ions $N_{Li}$ for AB- and AA stacked bilayer graphene. Local minima in $E_{int}$ correspond to $C_{14}LiC_{14}$ stoichiometry. Solid lines show the results of calculations where Coulomb interactions in the out-of-plane direction were truncated to avoid spurious long-range interactions (as appropriate for BLG) and dashed lines the results without truncation (corresponding to intercalation energy for bulk graphite). *Inset*: Entropy contribution vs Li density for AA and AB stacking, blue and black line, respectively. **(b-d)** Gibbs free energy vs lithium density at values of $\Delta\mu$ covering the range of its experimental uncertainty (see text). In experiment, $\Delta\mu$ was found to be 0.4±0.02 eV. Labels show $N$ for the corresponding $C_N LiC_N$ stoichiometries. Bright blue and black symbols correspond to Li ion configurations with the hexagonal symmetry and light blue/grey to non-hexagonal configurations with non-equidistant Li ion positions. The solid lines connecting data for the hexagonal symmetry are guides to the eye.

Using the above intercalation energy and incorporating entropy terms, the Gibbs free energy (relative to unintercalated state) at different Li densities is

$$\Delta G = \rho E_{int} - \rho \Delta\mu + k_B T \left[ \bar{\rho} \ln(\bar{\rho}) + (1-\bar{\rho}) \ln(1-\bar{\rho}) \right], \qquad (2)$$

where $\rho = N_{Li}/N_C$ gives the density of Li ions, $\bar{\rho} = \frac{\rho N_C}{N_{sites}}$ accounts for the number of available intercalation sites [14] and $\Delta\mu$ can be related to experimentally measured values of the pseudo-reference potential (Supplementary Fig. 7). In experiment $\Delta\mu = 0$ is equivalent to the pseudo-reference potential, $V_{ref}$, corresponding to the start of Li ions entry into the bilayer, i.e., to the sharp peak in $\rho_{xx}$. Accordingly, $\Delta\mu$ in eq. (2) corresponds to the change of $V_{\text{ref}}$ resulting from Li intercalation, $\Delta\mu = e\Delta V_{\text{ref}} = 0.4 \pm 0.02$ eV, see Supplementary Fig. 7. For the temperature fixed at the experimental value $T = 330$ K and $\Delta\mu = 0.4$ eV, the evolution of the Gibbs free energy as a function of Li density (ratio of the number of Li ions to the number of carbon atoms) is shown in Fig. 6 in the main text. To accommodate the experimental range of $\Delta V_{\text{ref}}$ obtained in different intercalation cycles, Supplementary Fig. 8b-c shows the DFT results for $\Delta\mu = 0.35, 0.4$ and $0.45$ eV, with the global minimum in the Gibbs energy remaining at $N = 14$ ($C_{14}LiC_{14}$ composition).

At low Li densities both the intercalation energy $E_{\text{int}}$ and the Gibbs energy $\Delta G$ are lower for AB stacking, while at higher $N_{\text{Li}}/N_{\text{C}}$ both energies are lower for the AA stacked bilayer. It is notable, however, that for the intercalation energy the crossover between AB and AA stacking occurs at a significantly higher Li to carbon ratio, $C_{32}LiC_{32}$ versus $C_{54}LiC_{54}$ for the Gibbs free energy, compare Supplementary Figs 8a and 8b-c. This is the consequence of a large difference in entropy contributions for the two bilayer stackings, because there are twice the number of available intercalation sites in AB-stacked BLG, compared to AA. The smaller number of available intercalation sites in AA-BLG saturates at a lower Li density and therefore favors AB over AA at low densities, see inset if Supplementary Fig. 8a. At $T = 330$ K, this effect is responsible for pushing the predicted AB-to-AA restacking transition ($\Delta G_{\text{AA}} = \Delta G_{\text{AB}}$) from $C_{32}LiC_{32}$ to approximately $C_{54}LiC_{54}$ (see Supplementary Fig. 8b-d). This result is in agreement with reported experimental observations of a large entropy contribution at low Li densities in intercalated graphite [15].

Both the intercalation energy $E_{\text{int}}$ and the Gibbs energy $\Delta G_{\text{AA}}$ show minima for the $\sqrt{7} \times \sqrt{7}$ superlattice of Li ions ($C_{14}LiC_{14}$ stoichiometry). In the latter case $C_{14}LiC_{14}$ corresponds to the thermodynamic equilibrium between the chemical potential of Li ions in the bilayer and in the electrolyte, $\frac{\partial G_{\text{AA}}}{\partial \rho} = 0$.

**2.2. Domain nucleation**

While the Li density corresponding to thermodynamic equilibrium is in excellent agreement with experiment, the calculated Li density corresponding to the Gibbs energy crossover from AB and AA stacking appears to be notably lower, $C_{54}LiC_{54}$ as opposed to experimental $C_{42}LiC_{42}$ / $C_{38}LiC_{38}$ where the change of bilayer stacking is inferred from the transition from stage I/II to stage III/IV (Fig. 3a). This is due to the fact that restacking requires nucleation of AA domains and these have to be of a sufficient size in order to continue to grow over the whole sample, similar to the growth of nanoparticles but in 2D. To estimate the minimum size of nucleating AA domains in an initially AB-stacked bilayer graphene, we approximate the domain energy as the sum of an areal energy density $\sigma$ associated with restacking of Li-intercalated BLG (energy gain) and an elastic penalty per unit length of the AA-AB domain wall $\gamma$:

$$E(r) = -\sigma A(r) + \gamma C(r). \qquad (3)$$

Here A and C are the area and circumference of a circular domain, respectively, and the critical radius for the formation of an AA domain is $r_c = 2\gamma/\sigma$. Here $\sigma$ = 2.6 meV/Å$^2$ is the calculated Gibbs energy difference for AB and AA stacking of the Li intercalated bilayer. For the line tension we could assume $\gamma \approx 0.1$ eV/Å as found in literature for AB/BA shear domain walls [16,17]. However, a circular domain wall between AA- and AB-stacked regions requires a constant Burgers vector at the boundary, which leads to significant bending of the wall and creates hydrostatic strain, in addition to the pure shear strain of a simple screw dislocation analyzed in refs [16,17]. To account for this, we have calculated the elastic energy explicitly in LAMMPS [18] using the AIREBO potential for intralayer interactions [19] and DRIP for interlayer adhesion [20]. Fixed AA domains were created inside a graphene bilayer, while all atoms outside of these fixed regions were allowed to relax, after which the total elastic energy was calculated as the difference between individual deformed monolayers and perfect graphene. The calculated line tension ($\gamma$ = 0.16 eV/Å) for a 20 nm AA region is notably higher than that of a perfect shear domain wall. To estimate the critical domain size, we have compared the adhesion and elastic energies as a function of domain radius as shown in Supplementary Fig. 9. This yielded $r_c \approx 12$ nm corresponding to ~480 Li ions per domain.

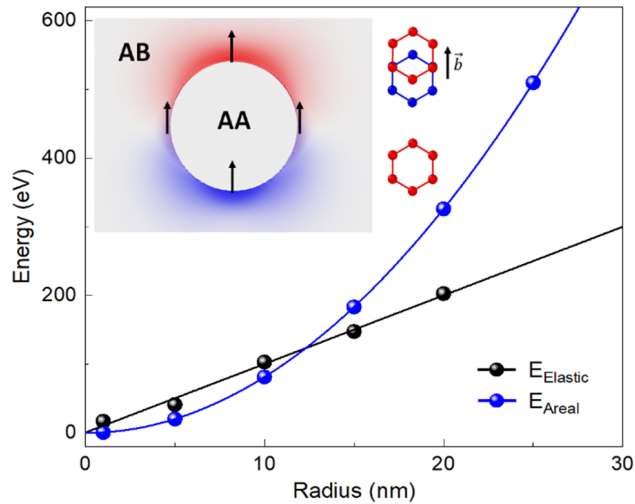

**Supplementary Figure 9 | DFT energy contributions to formation of an AA domain in an initially AB-stacked bilayer intercalated with Li.** Black symbols: elastic energy as a function of the domain radius. Blue symbols: areal stacking energy. *Inset*: Strain distribution around an AA-stacked domain. Hydrostatic tension and compression alternate at different sides of the domain, resulting from the constant Burgers vector at the surrounding domain walls as shown to the right.

## 2.3. Lithium ion diffusion

Our experiment suggests significant differences in Li ion diffusivity in AB- and AA stacked bilayers. These differences can be understood by considering the energy of a Li ion as a function of lateral displacement within the BLG. Supplementary Fig. 10a,b shows the energy maps for a Li ion diffusing in AA- and AB-stacked bilayers. These are computed, neglecting relaxation, for the Li ion position in the vertical direction halfway between two rigid graphene layers. Supplementary Fig. 10c,d shows corresponding energy barriers for optimal diffusion paths by one lattice constant, which involves a single jump between adjacent sites for AA stacking and two jumps between energetically equivalent sites for AB stacking. Notably, for AA stacking this barrier is higher due to more significant overlap of Li and C valence electrons.

The computed DFT activation barriers can be related to diffusivity $D$ [21,22]:

$$D = \frac{1}{2d}\Gamma_T \alpha^2,$$

where $\alpha$ is the jumping distance ($\alpha = a$ and $\alpha = a/\sqrt{3}$ for Li ions in AA and AB-stacked BLG, respectively), $d = 2$ is the dimensionality of the diffusion process, and $\Gamma_T$ is the total frequency of the jumps between adjacent sites. The total frequency is $\Gamma_T = \sum_n \omega_n$, where $n$ is the number of adjacent sites ($n = 3$ for AA- and $n = 6$ for AB stacking), and $\omega_n$ is the frequency of a single inter-site diffusion event. The latter is calculated from DFT energy barriers using the Arrhenius equation

$$\omega_n = \nu \exp(-\beta E_{a,n}),$$

with $\beta = 1/k_B T$, $\nu \sim 10^{13}$ Hz the attempt frequency of a single diffusive jump [21], and $E_{a,n}$ the DFT energy barrier. Our DFT calculations yield the energy barriers $E = 0.28$ and $0.07$ eV for Li diffusion in AA- and AB-stacked BLG, respectively, with corresponding diffusion constants $D \approx 2.4 \times 10^{-7}$ and $2.6 \times 10^{-4}$ cm$^2 \cdot$s$^{-1}$. These values suggest significantly faster ion diffusion in AB-stacked bilayer, in agreement with experiment.

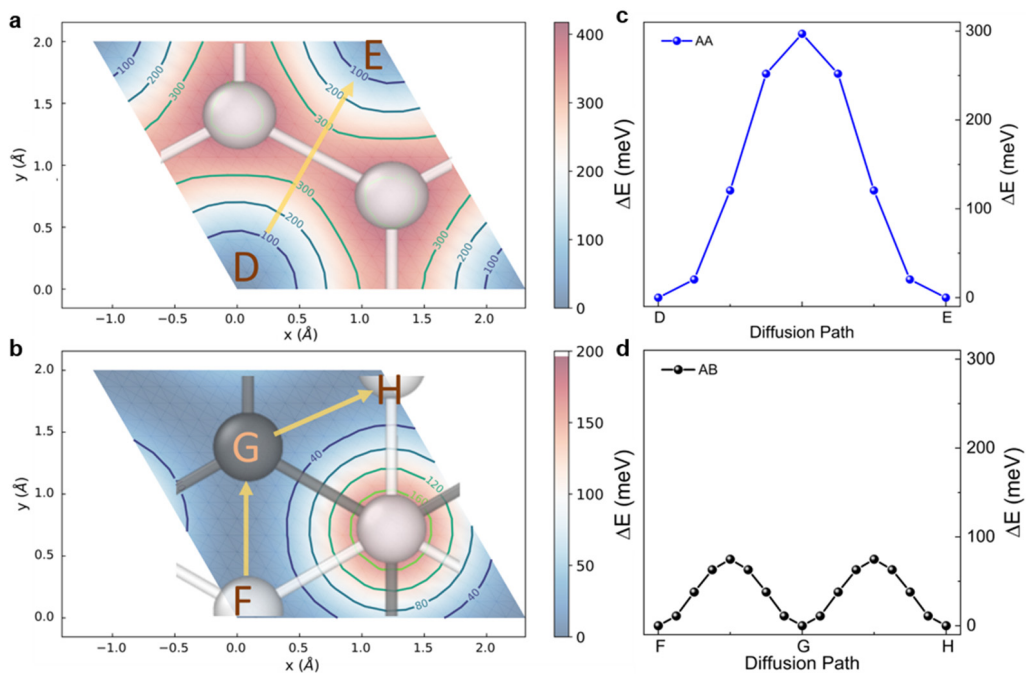

**Supplementary Figure 10 | Nudged elastic band (NEB) barriers for a Li atom diffusing in AA- & AB-stacked bilayer graphene. (a,b)** Energy landscape for Li-intercalated AA- and AB- stacked bilayers. The arrows show optimal diffusion paths for Li ion jumps by one lattice constant via saddle points in the energy maps. **(c,d)** Corresponding activation energies. In AB-stacked bilayer, Li ions can entirely avoid the large maximum associated with the interlayer C-C dimer.

### 2.4. Screening of interionic interactions.

The importance of electrostatic repulsion between intercalated ions in layered materials, where ions intercalate into interlayer galleries, has long been appreciated, notably for graphite [25-28], MXenes [29] and TiS$_2$ [30]. The authors of refs. [38, 41] (main text) used a simple model of point charges interacting through screened electrostatic interactions to argue that screening has a strong effect on attainable intercalation densities. This effect is especially important for intercalation of ions into BLG, where the same planar density of Li ions per carbon layer (C$_x$LiC$_x$) as in bulk graphite (LiC$_x$) results in less Li-dense configurations overall, simply because the same number of Li ions are shared by the two adjacent layers in BLG but not in bulk graphite. As the result, the host carbon atoms are less strongly doped, because the amount of charge lost by one Li ion to the surrounding graphene layers is approximately the same in both cases. Accordingly, the density of charge carriers at the Fermi level in Li-intercalated BLG is lower, which in turn leads to relatively ineffective screening of the interionic repulsion.

Our DFT calculations incorporate the effect of screening naturally and allow numerical estimation of its impact. We find a very strong effect of the poorly screened Coulomb repulsion on ionic mobility, as shown in Supplementary Fig. 11 through comparison of off-stoichiometric arrangements of Li ions in a repeated $\sqrt{7} \times \sqrt{7}$ superlattice. It demonstrates a significantly higher energetic penalty in the bilayer case, in agreement with the qualitative arguments above.

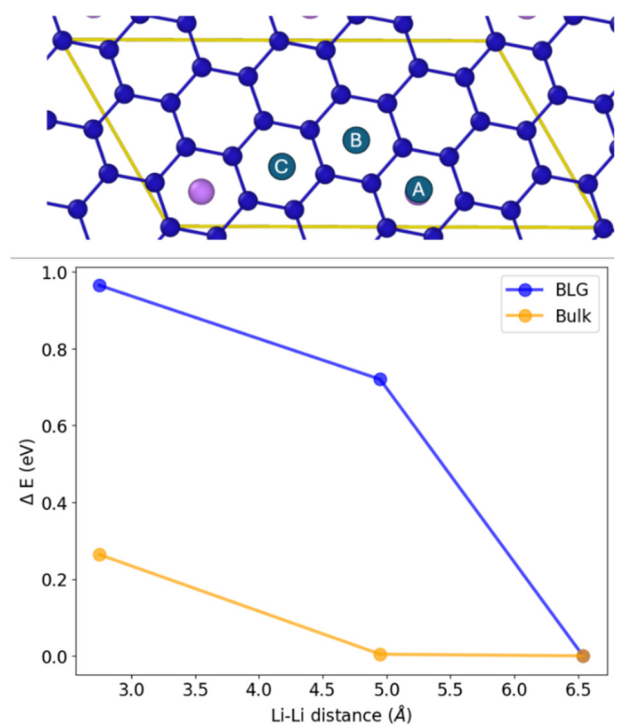

**Supplementary Figure 11 | Comparison of the energy penalty for off-stoichiometric arrangements of Li-ions in bulk graphite and bilayer graphene.** *Top panel*: Three distinct structures considered in the calculations. Each structure contains two Li ions but the distances between the two ions are different in each case. For structure (A) the position of the second Li ion corresponds to the experimentally found stoichiometry, with Li-Li distance ≈ 6.5 Å. For (B) and (C) the inter-ionic distance is artificially decreased in order to quantify the Coulomb repulsion. *Bottom panel*: Increase in the interaction energy (referenced to the experimental stoichiometry) as a function of inter-ionic distance for bulk graphite (yellow symbols) and for BLG (blue). The energy penalty is markedly higher for the bilayer.

## Supplementary References